\documentclass[prd,twocolumn,showpacs,nofootinbib]{revtex4}
\usepackage{amsmath}
\usepackage{bm}
\usepackage{graphicx}
\usepackage{graphpap}
\def\Journal#1#2#3#4#5#6{(#5) ``#6'' {#1} {\bf #2} #3#4}
\def\sign{{\rm sign}}

\font\twelvemsb=msbm10 scaled 1200 
\def\Bbb#1{\hbox {\twelvemsb#1}}

\newtheorem{theorem}{Theorem} 
\newtheorem{proposition}{Proposition} 
\newtheorem{lemma}{Lemma} 
\newtheorem{corollary}{Corollary}

\newtheorem{defi}{Definition}

\def\CQG{\em Class. Quantum Grav.}

\def\PRD{\em Phys. Rev. D }
\def\GRG{\em Gen. Rel. Grav.}
\def\IJT{\em Int. J. Theor. Phys.}

\def\JMP{\em J. Math. Phys.}

\def\PRL{\em Phys. Rev. Lett.}

\def\PREP{\em Phys. Rep.}

\def\PLB{\em Phys. Lett. B } 

\def\NC{\em Nuovo Cimento} 
\def\EPL{\em Eurphys. Lett.} 
\def\JHEP{\em JHEP}
\def\JCAP{\em J. Cosm. Astropart. Phys.}
\def\ANN{\em Ann. Phys.}
\def\IJMPA{\em Int. J. Mod. Phys. A}

\def\d{\mbox{d}} 
\def\g{{\rm g}}

\def\S{\Sigma}
\def\proj{P}

\def\otheta{\underline{\theta}}

\def\hatell{\hat{\ell}}

\def\tension{\Lambda}
\def\lambdafour{\Lambda_{4}}

\def\nany{N}
\def\normal{n}

\newcommand{\M}{\mathcal{M}}

\newcommand{\V}{\mathcal{V}}

\newcommand{\N}{\mathcal{N}}

\def\BA{\tilde{A}}
\def\BB{\tilde{B}}
\def\BC{\tilde{C}}

\def\Br{\tilde{r}}

\def\Blambda{\tilde{\lambda}}
\def\Bt{{\tilde{t}\,}}
\def\Bphi{\tilde{\phi}}

\def\eqq{\stackrel{\S}{=}}

\def\Part{\partial_t}
\def\Partt{\partial_{\Bt}}
\def\Parr{\partial_r}
\def\Parrr{\partial_{\Br}}

\def\und{\underline}
\def\tt{\mathcal{G}}
\def\hh{\mathcal{H}}

\def\ttt{\tau}
\def\ttm{{\und{T}^\M}}
\def\ff{N}
 
\def\proof{\noindent{\em Proof.\/}\hspace{1mm}} 
\def\fin{\hfill \rule{2.5mm}{2.5mm}\\ \vspace{0mm}} 
\def\finn{\hfill \rule{2.5mm}{2.5mm}}

\begin{document} 
\title{Lorentzian and signature changing branes} 
\author{Marc Mars}
\affiliation{\'Area de F\'{\i}sica Te\'orica, Universidad de Salamanca,
Plaza de la Merced s/n, 37008 Salamanca, Spain}
\author{Jos\'e M. M. Senovilla}
\author{Ra\"ul Vera}
\affiliation{F\'{\i}sica Te\'orica,  
Universidad del Pa\'{\i}s Vasco, Apartado 644, 48080 Bilbao, Spain}
%\thanks{Also at Laboratori de F\'{\i}sica Matem\`atica, 
%Societat Catalana de F\'{\i}sica, IEC, Barcelona, Spain.}  

%\date{\today}  
\begin{abstract}
General hypersurface layers are considered in order to describe brane-worlds and shell cosmologies. No restriction is placed on the causal character of the hypersurface which may thus have internal changes of signature. Strengthening the results in our
previous letter \cite{letter}, we confirm that a good, regular and consistent description of signature change is achieved in these brane/shells scenarios, while keeping the hypersurface and the bulk completely regular. Our formalism allows for a unified description of the traditional timelike branes/shells together with the signature-changing, or pure null, ones. This allows for a detailed comparison of the results in both situations. An application to the case of hypersurface layers in static bulks is presented, leading to the general Robertson-Walker geometry on the layer ---with a possible signature change. Explicit examples on anti de Sitter bulks are then studied. The permitted behaviours in different settings ($Z_{2}$-mirror branes, asymmetric shells, signature-changing branes) are analysed in detail. We show in particular that (i) in asymmetric shells there is an upper bound for the energy density, and (ii) that the energy density within the brane vanishes when approaching a change of signature. The description of a signature change as a `singularity' seen from within the brane is considered. We also find new relations between the fundamental constants in the brane/shell, its tension, and the cosmological and gravitational constants of the bulk, independently of the existence or not of a change of signature.
\end{abstract} 

\maketitle

\section{Introduction}
In a previous letter \cite{letter}, we explicitly showed that brane-world models \cite{arkani,r-sI,r-sII} constitute a natural  
scenario for the classical {\em regular} description of a change of
signature in the physical spacetime. This can also be said in general of
any higher-dimensional theory (e.g. \cite{barcelo,hig-dim} and references
therein) admitting domain walls or (hyper)-surface layers. The same idea, together with the possibility of topology change, was later advocated in \cite{GI} by constructing explicit solutions of an action given by the area/volume functional of the brane.

In this paper we want to elaborate on the ideas put forward in \cite{letter}, but keeping the full generality so that we can also recover the traditional results usually derived for pure timelike branes or shells, see \cite{AABS,SMS,gogb1,derdol,Carter,gergely1,roy_review,AS}. We also want to present detailed proofs of several interesting results merely announced in \cite{letter}.

In geometrical terms, branes or shells are submanifolds of a higher
dimensional spacetime called the bulk. The bulk metric
is differentiable everywhere except on the brane where it is 
only continuous. The jump in the derivatives of the bulk metric is related
to the part of the {\em distributional} energy-momentum tensor with support on the brane.
Traditionally, branes have been assumed to be timelike
submanifolds so that the induced geometry is Lorentzian and the brane can
describe the four-dimensional spacetime where we live. In this case, the precise relation between the jumps in the metric derivatives and the energy-momentum quantities on the brane is given by the so-called Israel conditions \cite{Israel66}. 

However, the timelike assumption is too strong: on physical grounds,
it is enough that there is a region of the brane where it is timelike.
A priori, there is no physical or mathematical obstruction to the existence of
{\it completely regular} branes ---in totally regular bulks---which change its causal character.
%from (say) spacelike to timelike, or which are partly null,  
%or even more complicated possibilities. 
As a trivial example, consider a circumference centred
at the origin of two-dimensional Minkowski spacetime: this has two spacelike regions, two timelike ones, and four points where it is null.
Therefore, in these higher-dimensional scenarios with branes or shells,
the study of signature change becomes the geometrical analysis of embedded  
submanifolds in the bulk: a well-posed mathematical problem.
%without pathologies.  
%Notice that the signature in the bulk is left unchanged, so that 
%our work differs significantly from other recent studies  
%\cite{bulk}. 

As we shall see, a very interesting property of this type of signature-changing branes is that, even though the change of signature may appear as a dramatical event  
when seen from {\em within the brane} ---specially if the scientists
living there believe that their Universe is Lorentzian everywhere---,
both the bulk and the brane can be totally smooth. As a matter of fact,
we shall see that the change of signature occurs at a region in the brane
which might be {\em interpreted} as a curvature singularity by those
scientists. Of course, this opens the door to explain, or avoid, 
the classical singularities of General Relativity \cite{sing,sing2}.
For instance, a past   
big-bang singularity may be replaced by a signature-changing set leading to an
Euclidean region prior to the birth of time, or
a neighbourhood of the singularity inside a black hole by a kind of
Euclidean core. Explicit examples of these situations
were actually built in \cite{letter,HH}, see also \cite{GI}.
This has
natural and obvious links with the no-boundary proposal \cite{noboundaryprop}
for the prescription of the wave function of the Universe
in quantum cosmology, and with similar ideas of quantum
tunneling \cite{Sak,vilenkintunnel} from `nothing' or from instantons. 
%In general, every process which can be studied by  
%resorting to the ``imaginary time'', e.g. \cite{noboundaryprop}, 
%can be also analyzed by means of change of the signature. 
%All these possibilities could be naturally considered in our proposal. 

From a classical standpoint, changes of signature were treated
in the literature mainly from the inner point of view, by just considering a manifold with a metric which becomes degenerate somewhere and changes signature, see \cite{chsig,chsig2,chsig3,ellissig} and references therein. There was much debate on whether the transition between the Euclidean and the Lorentzian regions should occur smoothly or with a jump, see \cite{DEH} and references therein. Both cases
can be treated in the brane scenario that we proposed in \cite{letter}.
The most natural case, though, is when the brane or shell is differentiable
and the signature change is therefore smooth \footnote{By
allowing continuous
piecewise differentiable hypersurfaces, 
we could also describe discontinuous changes of
signature. This would require a detailed knowledge of the
matching conditions within a submanifold which, itself, is a matching
hypersurface between the two bulk subregions. The mathematical tools needed for that purpose have only appeared recently in the literature \cite{taylor}.
Incidentally, a similar comment holds for standard timelike branes whenever the
energy-momentum tensor within the brane has jumps, as for instance on
the surface of a star. A proper mathematical description
of such a situation would require the results in \cite{taylor}. Let us remark that
several papers have actually dealt with stars on the brane, e.g.
\cite{Germani01,Dadhich02}.  However, in those papers the whole
description is made from within the brane, with standard matching
conditions across the surface of the star. It would be interesting to
perform a full bulk description and compare the results.}.
Some explicit signature changing solutions of the field equations for
scalar field sources \cite{DT,DR} and for spinor fields \cite{VJS,VS}
have been found, as well as in higher dimensional cases with compact
extra dimensions \cite{VJS2}, or for the spherically symmetric case
\cite{DDT}. In any case, the differences between our approach
and those intrinsic treatments are radical,
as we have a bulk structure available which defines inherited regular
structures on the brane. As a matter of fact, we can even prove
that some {\it ad hoc} assumptions in
\cite{chsig,chsig2,chsig3,ellissig,DEH} become {\it necessary conditions}
in our setting,
precisely because of this bulk structure, see Sect. \ref{restrictions}.
 
In order to describe signature-changing branes we need to consider hypersurfaces without a fixed causal character, so that they may have timelike, spacelike and null portions. There are some obvious technical difficulties when dealing with hypersurfaces of such an arbitrary causal character. For instance, the first fundamental form is degenerate somewhere, and also the second fundamental form is no longer {\em extrinsic}  everywhere---it is actually intrinsic at null points. 
This leads to the most important difficulty:
the usual matching conditions are no longer valid for hypersurfaces with changing causal character. In particular, the Israel formulas \cite{Israel66} are not suitable to describe the energy-momentum on the shell or brane, and 
the appropriate generalization must be used. Probably this has been the reason behind the lack of studies on signature-changing branes prior to \cite{letter}, and also of some misunderstandings \footnote{In \cite{GGI} the traditional Israel formula (\ref{Israel}) is applied to a $Z_{2}$-symmetric brane, and then the brane is claimed to undergo a change of signature. Nevertheless, for a signature-changing brane the Israel conditions (\ref{Israel}) are not valid, and the appropriate generalised formula (\ref{eq:tt}) should have been used. Observe that, in fact, a $Z_{2}$-symmetric brane can never undergo a signature change, according to our general result Corollary \ref{res:2}.} in the interesting recent work \cite{GGI}.
Fortunately, the required generalization was already developed in \cite{MarsSenovilla93} 
in four spacetime dimensions.  
The results carry over to any dimension with no essential change 
and can therefore be used to study signature-changing branes.
A self-consistent summary of the required
results from \cite{MarsSenovilla93} is
presented in sections \ref{sec:basic} and \ref{genIsrael}.

When dealing with changes of signature, there always remains the important unsolved question of which physical mechanisms may produce, or induce, them. Several speculative possibilities have appeared in the literature, such as large time fluctuations
\cite{Sha}, tachyon condensation and S-branes \cite{HHW,HHNW,KP,KM}, dynamical stabilization of extra dimensions by means of scalar fields \cite{JAS}, or emergent spacetimes in Bose-Einstein condensates \cite{WWV}. In particular, one should be able to device a physical process stimulating the signature change on a brane of an otherwise innocuous bulk. As far as we are aware, this is an important and fully open question, which we shall not address in this paper.

\subsection{Plan of the paper and summary}
The basics of gluing and how to construct branes or shells
by pasting together two spacetimes with boundary is presented in
section \ref{sec:basic}. When the boundaries have a non-constant signature,
one needs the results of \cite{ClarkeDray87,MarsSenovilla93} to perform the
matching correctly. These are described, and corrected, in that section.
In particular, we also correct an erroneous statement in our
letter \cite{letter}. In this section we also prove rigorously that
signature-changing branes or shells are not compatible with the
$Z_{2}$-mirror symmetry assumed many times in braneworlds.

The generalised Israel formula is then presented and briefly explained
in Section \ref{genIsrael}. Section \ref{fieldeqs} is devoted to the
field equations on the bulk and their consequences on the brane or shell.
In particular, we prove some statements announced in \cite{letter}: there cannot be umbilical hypersurfaces changing signature, and the brane tension cannot be constant in signature-changing branes.

Readers familiar with these matters may skip the mentioned sections and
go directly to the more interesting physical results discussed afterwards.

Sections \ref{static} and \ref{AdSect} deal with the explicit construction of (signature-changing or not) branes and shells in static and spherically, plane, or hyperbolically symmetric bulks. We prove that, by gluing two such spacetimes across any hypersurface preserving the spatial symmetries, a brane or a shell is obtained which has the general Robertson-Walker line-element, with a possible change of signature, as first fundamental form. The physical quantities of these branes are then computed in general. Section  \ref{AdSect} specializes these results to the case of two anti de Sitter spacetimes, which produces a bulk with two, different in principle, cosmological constants. The different possibilities are then analyzed in detail, and we recover all previous results on $Z_{2}$-symmetric branes and asymmetric shells. 

We also derive the corresponding new results for the signature-changing branes or shells, and we prove that the boundary of the Lorentzian part of the brane is part of a signature-changing set which is completely regular. We further show that the energy density of the matter fields {\em vanishes} when approaching this set. The possible interpretation of this set as a curvature singularity for observers living within the Lorentzian part of the brane is carefully considered. Finally, old {\em and new} relations between the fundamental constants in the brane and the parameters of the bulk are derived in some physically motivated limits.

Throughout the paper, we will use units with the speed of light $c=1$.

\section{Signature changing hypersurfaces: basic
properties and brane construction by gluing}
\label{sec:basic}
In general, branes are submanifolds in a higher dimensional
spacetime $(\M,g)$, which is called the bulk, with $g$ being a Lorentzian metric of 
signature $(-,+,\dots ,+)$. Such objects had been traditionally known (specially for the case of co-dimension one) as thin layers or domain walls. The typical branes have a constant causal character, usually timelike. However, the purpose of this paper is to analyze the possibility of having more general branes such that their causal character may change from point to point. Thus, 
{\em signature changing branes} are submanifolds of changing causal
character in $(\M,g)$. In this paper we will present a unified
formalism which allows to deal with general types of branes, signature-changing and signature-constant ones, at the same time.

In order to have a topological defect such as a co-dimension one brane,
{\em and} to have well-defined Einstein's field equations on the entire bulk, the metric
$g$ needs to be at least of class $C^2$ everywhere on $\M$ except on the brane $\S$, where
it should only be continuous ---in a suitable coordinate system.
Thus, the differentiability
of the manifold $\M$ must be at least $C^3$, which we will assume from
now on.
$C^3$ manifolds with $C^2$ Lorentzian metrics will be
called $C^2$ spacetimes from now on.
%\mnote{\marc Necessitem aquesta definicio al Teorema 1}
The brane $\S$ is a hypersurface and hence it inherits a first fundamental
form $h$ which must also be at least $C^2$ in order to admit gravitational field equations
within the brane \footnote{If objects like stars, with discontinuous
energy-momentum tensors, are present on the brane, then the $C^2$
differentiability of $h$ holds only outside the separating surfaces.}.
Consequently, as a submanifold $\S$ must be at least $C^3$ too.
%\mnote{\marc Aqui hi havia un paragraf que quedava fora de lloc. Just abans ja estavem parlant
%de branes com a ``topological defects'' i fent servir matching.
%Si es vol mantenir, hauria d'anar abans, pero jo el treuria completament.}
%The case most extensively treated in the
%literature is that of branes as hypersurfaces  dividing
%$\M$ into two submanifolds-with-boundary, $\M^{+}$ and $\M^-$. This is a global
%requirement which need not always hold. Nevertheless, this is always
%satisfied for branes built by gluing two spacetimes with boundary, which is
%one of the most popular constructions in the standard literature.
%In this paper, we will also use this type of construction, which implies, in particular,
%that the brane must have co-dimension one.
%Let $\S$ be a hypersurface on $\M$, our candidate for a brane. 
Let $\bm{\nany}$ be a normal one-form of $\S$, i.e. a non-zero one-form 
satisfying
$$
\bm{\nany}(\vec v)=0 \hspace{5mm} \forall\,  \vec v\in T_p\S , \hspace{2mm} \forall p\in \S
$$
so that it annihilates all vectors {\em tangent} to $\S$.
Raising the index of $\bm{\nany}$ we obtain a vector field $\vec N$ which
may still be called ``normal vector field'', but which is not necessarily
transverse to $\S$ everywhere.
From our assumptions,
$\bm{\nany}$ and $\vec N$ are differentiable fields. Observe that, if we want to allow for signature changes on the brane, $\bm{\nany}$ cannot  be globally normalized as it is null somewhere. Thus, $\bm{\nany}$ is
defined only up to rescaling $\bm{\nany} \rightarrow A \bm{\nany}$, where $A$ is a nowhere
vanishing $C^2$ function on $\S$. This ``normalization'' freedom plays an important
role in the physics of the brane and needs to be kept in mind. Since the bulk metric is
continuous across $\S$, the norm (Greek lower-case indices run from
$0, \cdots, n-1$, where $n$ is the dimension of the bulk)
$$\bm{\nany}(\vec N)\equiv (\bm{\nany},\bm{\nany} ) \equiv  g^{\mu\nu} \nany_{\mu} \nany_{\nu}$$ 
is well-defined on $\S$ and of class $C^2$.
For the signature of the brane to change,  the set of 
points where $\S$ is null must be non-empty.
Thus, to fix ideas and notation, we put forward the following
\begin{defi}
Let $\S_E \subset \S$, $\S_0\subset \S$, and $\S_L \subset \S$ be the subsets where 
the hypersurface $\S$ is spacelike, null and timelike, respectively. Equivalently,
\begin{eqnarray*}
\S_E\equiv \{p\in \S : (\bm{\nany},\bm{\nany})|_p < 0\},\\
\S_0\equiv \{p\in \S : (\bm{\nany},\bm{\nany})|_p=0\},\\
\S_L\equiv \{p\in \S : (\bm{\nany},\bm{\nany})|_p > 0\}.
\end{eqnarray*}
Accordingly, the induced metric $h$ is positive definite
at $\S_E$, Lorentzian at $\S_L$ and degenerate at $\S_0$. Then,
$\S_E$ is called the {\em Euclidean phase} of the brane,
$\S_L$ its {\em Lorentzian phase}, and $\S_0$ its {\em null phase}.
Finally, the set 
$$
S\equiv \left(\overline{\S_L}\cap \S_0\right)\cup \left(\overline{\S_E}\cap \S_0\right)
$$
is called the {\em signature-changing set} of $\S$.  
\end{defi}
By definition $\S_0$ is a closed 
%defined as the set where $(\bm{\nany},\bm{\nany})$ vanishes, this is a closed 
subset
of $\S$. Also by definition we have $S\subset \S_0$.
The case when $\S_0$ has empty interior is characterized by $S=\S_0$,
and will be one of the important cases in our analysis. Note also that $\bm{\nany}(\vec N)|_{\S_0}=0$ so that $\vec N$ is actually tangent to $\S$ on $\S_0$, see \cite{MarsSenovilla93}.

We will implicitly assume that $\S_L$ is non-empty so that we have at least one region
where the brane is timelike and therefore able to describe a real (Lorentzian) world.
Notice, though, that it is still possible that both $S$ and $\S_L$ are non-empty
while $\S_E = \emptyset$. In fact it is even possible to have branes
which are timelike everywhere except at a single point, where it is null.
A simple example is given by the hypersurface 
$$ \{ x=a(t)\cos \theta ,y = b \sin \theta, t \}$$ 
in 3-dimensional Minkowski spacetime with Cartesian coordinates $\{t,x,y\}$,
where $b>0$ is a constant and $a(t)$ is a positive function whose
derivative satisfies $|\dot{a}| \leq 1$ with equality at one single value.
Similarly, branes which are spacelike everywhere except
for a single point are possible, as well as {\em null branes} so that $\S_L=\S_E=\emptyset$. Most of these situations do not
truly describe a signature changing brane, or at least not the one we usually have in mind, which require that both $\S_L$ and $\S_E$ ---and therefore $S$ too---  are non-empty.
In this situation it is obvious that $S$ cannot consist of a finite
number of points. Even though our main goal in this paper are proper
signature-changing branes, {\em all} mentioned cases are included and can
be treated within our formalism. In the explicit examples, however, we will
mainly deal with proper signature-changing branes with $\S_0=S$, i.e.  such
that there is no open set where the hypersurface $\S$ is null.

\subsection{Restrictions on the signature-changing set $S$}
\label{restrictions}
Even if one assumed that $\S_0$ has empty interior there remains a lot of freedom on the structure of $S$. In a general setting, not necessarily of brane type, the signature changing set $S$ 
may have many different structures. Nevertheless, this is no longer true
in a brane-in-bulk setting, which is a desirable outcome, because conditions
on $S$ which are typically assumed ad hoc become predictions in this
scenario. As a matter of fact, in our letter \cite{letter} we claimed that
one of the advantages of studying signature change within the brane scenario
is that the structure of $S$ becomes restricted. While this general claim
remains true, see Lemma \ref{rest} below, the 
specific result on the structure of $S$ presented in \cite{letter} is unfortunately
false. We are grateful to E. Aguirre-Dab\'an and J. Lafuente-L\'opez \cite{AL}
for pointing out that Result 1.1 in \cite{letter} is not correct.
Let us describe this in detail.  

Result 1.1 in \cite{letter} states
that in the brane scenario, changes of signature occur at a single
``instant of time''. In other words, that $S$ is a spacelike
$(n-2)$-submanifold of the bulk. If we define, as usual, the radical
of a degenerate metric $h$ as the set of vectors $\vec{V}$ satisfying
$h(\vec{V},\cdot)=0$, the claim above amounts to saying that the first
fundamental form $h$ of $\S$ at $p \in S$ has a transverse radical
(i.e. that the radical is nowhere tangent to $S$). A detailed study
of signature changes with {\it tangent} radical
(i.e.  such that the degeneration vectors are tangent to the signature
changing set $S$) has been performed in \cite{AguirreLafuente}. 
This analysis was done in full generality,
without assuming that the signature changing
space $(\S,h)$ is a brane within a bulk. From these general results,
explicit examples of signature changes for branes with tangent radical
may be derived \cite{AL}. One such example is as follows. For signature changes with
$\S_0=S$ and tangent radical 
%(see \cite{AguirreLafuente} for the precise definitions),
there exists \cite{AguirreLafuente} a coordinate system
$\{y, x^{i}, v  \}$ ($i,j=3, \cdots, n-1$)
%$n-1$ is the dimension of $\S$
on a neighbourhood of any point $p \in S$ such that $S : \{y =0 \}$
and the signature changing ``metric'' reads
\[
ds^2 |_{\S} = {dy}^2 +
y \left (g_2 dv + g_{i} dx^i \right )^2 + g_{ij} dx^i dx^j,
\]
where $g_2$, $g_i$ and $g_{ij}$ are differentiable functions of $(y,
x^i,v )$ such that $g_2(0,x^i,v)=1$ and $(g_{ij})$ is positive
definite. This tensor can be obtained as the first fundamental form
of the hypersurface $\S : \{t=0\}$ in an $n$-dimensional bulk
spacetime with metric
\[
ds^2 = y dt^2 + 2 k dt dv + dy^2 + y \left (k_2 dv + k_{i} dx^i \right )^2 +
k_{ij} dx^i dx^j,
\]
where $k$, $k_i$, $k_2$ and $k_{ij}$ are functions of $(t,y,x^i,v)$
satisfying $k_2 |_{t=0}=g_2$, $k_{i} |_{t=0}=g_{i}$, $k_{ij}
|_{t=0}=g_{ij}$ and $k$ is chosen so that $ds^2$ has Lorentzian
signature everywhere. To see an explicit example (in four
dimensions, for definiteness) consider the metric
\begin{equation}
ds^2 = f ( dt^2 + dv^2) + 2 \left ( \sqrt{1 + f^2} \right) dt dv
+ dy^2 + dz^2,
\label{example}
\end{equation}
which is a globally defined, smooth, Lorentzian metric on
$\Bbb{R}^4$ for any smooth choice of $f(t,v,y,z)$. Take $f=y$ and the brane 
defined by $\S : \{t=0\}$, which is
Lorentzian for $y < 0$ and Riemannian for $y > 0$. The signature
changing set $S\subset \S$ is defined by $\{ y = t = 0 \}$, which is clearly a
two-dimensional {\it null} surface, contradicting Result 1.1 in
\cite{letter}. In fact, the same example (\ref{example}) with a
different $f$ can be used to show that $S$ needs not even be a
differentiable submanifold and that branch points are allowed. Indeed,
taking $f = yz$, the signature changing set is located at $yz =0$,
which are two 2-planes intersecting at the branch line $(y=z=0, v \in
\Bbb{R})$. Thus, Result 1.1 in \cite{letter} does not hold and the structure of $S$
allows for much more freedom than claimed there.

Despite several efforts, the only restriction on the structure of $S$, and more generally on the properties of $\S_0$, that we have been able to derive from the brane setting is
\begin{lemma}
\label{rest}
At any point $p \in \S_0$ (and therefore at all points of $S$) of a co-dimension one brane $\S$, the induced metric $h$ of $\S$ has a {\em unique} degeneration direction given by $\vec{\nany} |_p$.
\end{lemma} 
\begin{proof}
At $p \in \S_0$ the normal vector $\vec{\nany} |_p$ is also tangent to
$\S$.  It obviously satisfies $g(\vec{\nany},\vec{v}) |_p = 0$ for any
vector $\vec{v} |_p \in T_p \S$, which clearly implies
$h(\vec{\nany}, \cdot ) |_p =0$,
so that $\vec{\nany} |_p$ is a degeneration vector. To
show uniqueness, let us take another degeneration vector
$\vec{w} |_p \in T_p \S$.
It follows $g(\vec{w}, \vec{w} ) |_p =0$ and
$g(\vec{\nany},\vec{w})|_p =0$, so that the two null vectors $\vec{\nany} |_p$
and $\vec{w} |_p$ must be parallel.
\finn
\end{proof}

\vspace{2mm}

This Lemma does indeed restrict the structure of
$S$ in the brane scenario because more general behaviours
can occur for arbitrary signature changes.
It may happen,
for instance, that the metric changes signature at a smooth hypersurface
where the radical is two-, or higher-, dimensional, or even spans the whole tangent space.
A simple example of the latter is given by the following $(0,2)$-tensor
in $\Bbb{R}^m$ ,
\[
ds^2 = - t dt^2 + t^2 \left (dx_1^2 + dx_2^2+ \cdots dx_{m-1}^2 \right).
\]

\subsection{Gluing}
Let us next discuss the standard procedure of how to build branes by gluing manifolds with boundary, and the possibility of actually constructing signature changing 
branes by that method. This is important as
most of the standard branes are constructed in this manner. 
%In fact, 
%locally around any point in the brane, the bulk can always be considered as
%the result of a matching between two regions across a hypersurface (globally
%this may fail as already discussed). 
However, for hypersurfaces with
changing causal character, the usual
matching conditions are no longer valid and an appropriate generalization
must be used. Fortunately, such a generalization was already developed in 
\cite{MarsSenovilla93} in four dimensions. These
results can be readily generalised to arbitrary dimension with no essential
change. Since we shall use this matching procedure extensively,
let us describe its essential features.

We start from two oriented $C^3$ $n$-dimensional
manifolds with boundary
$\M^{\pm}$, whose boundaries are $\S^{\pm}$. These manifolds are endowed
with $C^2$ Lorentzian metrics $g^{\pm}$. 
In order to join them across their boundaries we need to identify the
boundaries pointwise. This means, in
particular, that there must exist a one-to-one correspondence between
$\S^{+}$ and $\S^{-}$, which moreover must be a diffeomorphism in order
to preserve the differential structure. Both for conceptual
and operational
reasons, it is convenient to state this condition in the
following equivalent manner: there exists an
abstract $(n-1)$-dimensional $C^3$ manifold
$\S$ and two $C^3$ embeddings
\[
\Phi_{+} : \, \, \, \S \,  \longrightarrow  \, \, \, \M^{+} , 
~~~ \Phi_{-} : \, \, \, \S \,  \longrightarrow  \, \, \, \M^{-}, 
\]
which satisfy $\Phi_{+} \left (\S \right) = \S^{+}$ and
$\Phi_{-} \left (\S \right) = \S^{-}$. The identification
of the boundaries is then given by
$\Phi \equiv \Phi_{+} \circ\, \Phi_{-}^{-1} |_{\S^{-}}$. Under
these circumstances, and using standard techniques of differential
topology, it follows that the space $\M \equiv \M^{+} \cup \M^{-}$,
with the boundaries identified, can be
endowed with a differential structure \cite{Hirsch}
so that it becomes a manifold.
%, i.e. we already have a
%differential manifold which, however, has no metric yet.
Our  aim is to define a metric $g$ on $\M$ which is 
continuous everywhere, in particular across $\S$
(we shall often abuse notation and identify $\S^{+}$,
$\S^{-}$ and $\S$ when necessary), such that
$g$ coincides with the original $g^{\pm}$
in the interiors of $\M^{\pm}$, respectively. Demanding continuity is 
obviously sufficient for having a well-defined induced
metric on the brane. It turns out that continuity is in fact the
only possibility, as we discuss next. 

\subsubsection{Tangent space identification: riggings}
As pointed out by Clarke and Dray \cite{ClarkeDray87}, defining
a metric on $\M$ requires not only that we identify 
the points on the boundary but also that the tangent spaces are
properly identified. The differential map 
$\d \Phi$ 
fixes uniquely the way of identifying the
tangent vectors of the boundaries. Thus, if we want to define a continuous
metric on $\M$
we need to require at least that the first fundamental
forms of $\S^{+}$ and $\S^{-}$ coincide (via $\Phi$). In other words
\begin{equation}
h^{+} \equiv \Phi_{+}^{\star} (g^{+}) = \Phi_{-}^{\star} (g^{-})
\equiv h^{-},
\label{firstmatching}
\end{equation}
where $\Phi_{\pm}^{\star}$ denote the pull-backs of $\Phi_{\pm}$ and $h^{+}, h^-$
are the first fundamental forms of $\S$
as defined from $\M^{+}$ and $\M^{-}$, respectively.
Conditions (\ref{firstmatching}) are called preliminary matching conditions.
When they hold we write $h = h^{+} = h^{-}$. In local coordinates they
read as follows. Let $\{\xi^{a}\}$ ($a,b=1,\cdots, n-1$),
$\{x^{\mu}_{+}\}$ and $\{x^{\mu}_{-}\}$ be local coordinate systems on
$\S$, $\M^{+}$ and $\M^{-}$  respectively. Consider
the 
%holonomic 
basis vectors $\frac{\partial}{\partial \xi^a}$ and their
images by $\d \Phi_{\pm}$ 
\begin{equation}
\vec{e}_a^{\, \pm} \equiv \left [\d \Phi_{\pm} \left ( 
\frac{\partial}{\partial \xi^a} \right) 
\right ] =
\frac{\partial x^{\mu}_{\pm} (\xi)}{\partial \xi^a}\, \frac{\partial}{\partial x^{\mu}_{\pm}},
\label{e's}
\end{equation}
where the functions $x^{\mu}_{\pm} (\xi)$ define the embeddings
$\Phi_{\pm}$ in local
coordinates, i.e. 
$$\Phi_{\pm} : \xi^a \rightarrow x_{\pm}^{\mu} (\xi^a).$$
Obviously, 
$\{\vec{e}_a^{\, \pm}\}$ span the tangent planes of the hypersurfaces
$\S^{\pm}$ as embedded in $\M^{\pm}$. 
In terms of these objects, the preliminary matching conditions (\ref{firstmatching}) read
\begin{equation}
h^{+}_{ab} (\xi) = h^{-}_{ab} (\xi), \label{prelim}
\end{equation}
where
$$
h^{\pm}_{ab} (\xi) \equiv g^{\pm}_{\mu\nu} \left(x^{\pm}(\xi)\right) \, \frac{\partial x^{\mu}_{\pm} (\xi)}{\partial \xi^a}\, \frac{\partial x^{\nu}_{\pm} (\xi)}{\partial \xi^b}. 
$$
In order to complete the identification of the tangent spaces, we
only need to identify one transversal vector on $\S^{+}$ with
one transversal vector on $\S^{-}$. Then, the identification of
all tangent vectors follows by linearity. To that end,  let us choose
a $C^2$ vector field $\vec{\ell}_{+}$ on $\S^{+}$ which is
nowhere tangent to $\S^{+}$. The existence of such 
a vector field, sometimes called {\it rigging} \cite{MarsSenovilla93}, 
is a standard property of manifolds with boundary. Transversality means 
$\nany^{+}_{\mu}\ell^{\mu}_{+}\neq 0$ where  $\bm{\nany}^{+}$ is a normal one-form of $\S^{+}$
in $\M^{+}$. Furthermore, we choose $\vec{\ell}_{+}$ pointing
towards $\M^{+}$ everywhere; actually, since $\vec{\ell}_{+}$ is transversal
to  $\S^{+}$, it is sufficient to impose that
$\vec{\ell}_{+}$ points towards $\M^{+}$ at one point of $\S^+$. 
Of course, we could alternatively demand that $\vec{\ell}_{+}$ points outwards from
$\M^{+}$. This would induce obvious changes in the discussion below
with no essential new features.

Now we need to choose another $C^2$ rigging
$\vec{\ell}_{-}$ on $\S^{-}$.
Since we intend to identify $\vec{\ell}_{+}$ with $\vec{\ell}_{-}$ and get
a continuous metric, we must impose {\em at least} that their
norms and
scalar products
with arbitrary vectors in $T\S$ coincide. This amounts to requiring that
\begin{equation}
g^{+}_{\mu\nu}\ell_{+}^{\mu} \ell_{+}^\nu  \eqq
g^{-}_{\mu\nu}\ell_{-}^{\mu} \ell_{-}^\nu,
~~~
g^{+}_{\mu\nu} \ell_{+}^{\mu} {e_{a}^{+}}^{\nu} \eqq 
g^{-}_{\mu\nu} \ell_{-}^{\mu} {e_{a}^{-}}^{\nu}
\label{rigg} 
\end{equation}
where the symbol $\eqq$ stands for equality using the diffeomorphism
$\Phi$. Equations (\ref{rigg}) should be interpreted as
restrictions on $\vec{\ell}_{-}$ once $\vec{\ell}_{+}$ has been chosen,
or vice versa. These $n$ conditions are not sufficient to ensure a proper matching,
as the rigging $\vec{\ell}_{-}$ must also satisfy
the property of pointing outwards from $\M^{-}$ everywhere. This is necessary
because, after the identification, the vector
$\vec{\ell} \equiv
\vec{\ell}_{+} = \vec{\ell}_{-}$ points towards $\M^{+}$
(as $\vec{\ell}_{+}$
does). When viewed from the glued manifold 
$\M = \M^{+} \cup \M^{-}$,
this is equivalent to saying that $\vec{\ell}$ points
outwards from $\M^{-}$. 
%Thus, for consistency the vector $\vec{\ell}_{-}$
%must have this property to begin with. 

%The algebraic conditions (\ref{rigg})
%together the choice of orientation are necessary
%conditions for a matching producing a continuous metric. For shortness we  call
%such a matching a {\it continuous matching}. 
Two important questions arise:
(a) are these conditions on the riggings already sufficient for the
existence of a matching with continuous metric? and (b) do they
introduce any restrictions on the manifolds $\M_{\pm}$ to be matched?
In a remarkable paper \cite{ClarkeDray87}, 
Clarke and Dray addressed these questions for the case of 
constant-signature matching hypersurfaces.
%everywhere timelike,everywhere spacelike or everywhere null matching hypersurfaces. 
Their conclusion was that the answer is affirmative in both cases.
%i.e. that once two manifolds-with-boundary (with boundaries of
%constant signature) satisfy the preliminary matching conditions, they
%can be matched to produce a spacetime with continuous metric. To be
%more precise the statement in \cite{ClarkeDray87} is
%\begin{stat}
%Let $(\M^{\pm},g^{\pm})$ be two $n$-dimensional $C^2$ oriented
%spacetimes-with-boun\-da\-ry, with respective $C^3$ boundaries
%$\S^{\pm}$.  Assume that the first fundamental forms $h^{\pm}$ on
%$\S^{\pm}$ have constant signature (either Lorentzian, Euclidean
%or degenerate) and that the preliminary matching conditions (i.e. the
%existence of a diffeomorphism between $\S^{+}$ and $\S^{-}$
%which maps $h^{+}$ into $h^{-}$) are satisfied.
%Then, there exists
%a unique, maximal, $C^3$ differentiable
%structure on $\M = \M^{+} \cup \M^{-}$ (with their points on
%$\S^{+}$ and $\S^{-}$ identified), and a unique continuous
%metric $g$ which coincides with $g^{+}$ on $\M^{+}$ and with $g^{-}$
%on $\M^{-}$. 
%\label{CD1}
%\end{stat}
Unfortunately, this conclusion is not completely correct as stated, as we shall
see presently with examples. Let us discuss this.
%of manifolds with boundaries
%which satisfy all the conditions of Assertion \ref{CD1} but which {\it
%cannot} be joined continuously.

The proof  given by Clarke and Dray can
be divided in two parts. In the first one, question (a) above is
addressed and the authors try to prove that a pair of riggings
$\vec{\ell}_{\pm}$ satisfying (\ref{rigg}) with the correct
orientation does exist. In the second part, which corresponds to
question (b) above, the existence of a maximal atlas on $\M$ for which
the metric $g$ is continuous is shown, provided the preliminary
matching conditions hold and a pair of suitable riggings
$\vec{\ell}_{\pm}$ exist. This second part is correct and, in fact,
depends very weakly on the assumption of constant signature of the
matching hypersurface. A slight modification of the argument allows
one to show that the same result holds for spacetimes with boundaries
having varying causal character.
The first part of the proof, however, is not correct {\em for boundaries
having null points}, both in the constant null-signature case treated
in \cite{ClarkeDray87} or in its generalization to signature changing
boundaries. Thus, a correct reformulation of Clarke and Dray's result is
\begin{theorem} Let $(\M^{\pm},g^{\pm})$ be two $n$-dimensional $C^2$
oriented spacetimes-with-boun\-da\-ry, with respective $C^3$
boundaries $\S^{\pm}$ such the preliminary matching conditions
(\ref{prelim}) hold on $\S$.
Assume further that there exist transverse vector fields 
$\vec{\ell}_{\pm}$ on $\S^{\pm}$ satisfying the scalar product
conditions (\ref{rigg}) and such that $\vec{\ell}_{+}$ points towards
$\M^{+}$ and $\vec{\ell}_{-}$ points outwards from $\M^{-}$. 

Then,  there exists a unique, maximal, $C^3$ differentiable
structure on $\M = \M^{+} \cup \M^{-}$ (with their points on
$\S^{+}$ and $\S^{-}$ identified), and a unique continuous
metric $g$ which coincides with $g^{+}$ on $\M^{+}$ and with $g^{-}$
on $\M^{-}$. 

\label{theorem1}
\end{theorem}
{\bf Remark.} The hypothesis on the existence of the rigging is 
necessary only in the case of boundaries which have at least one point
of degeneration, i.e. $\S_0\neq \emptyset$. For everywhere spacelike or
everywhere timelike boundaries the unit normal vectors with appropriate
orientation fulfil all the requirements.

\subsubsection{On the existence of riggings for $\S$ with null points}
\label{subsub}
When $\S$ has null points existence of the appropriate riggings is not guaranteed, 
as we show next. We start with a Lemma stating that, 
at points where the hypersurface is non-null, the solution
of (\ref{rigg}) with the proper orientation is unique, if
it exists.
\begin{lemma}
\label{uniquenessNonNull}
Let $\M^{\pm}$ be two spacetimes-with-boundary satisfying the
preliminary matching conditions (\ref{prelim}). Let $\S^{-}$ be non-null at $p^{-} \in \S^{-}$ 
and set $p^{+} = \Phi(p^-)$.  Choose any
transverse vector field $\vec{\ell}_{+} |_{p^+}$ pointing towards $\M^{+}$.
Then there is at most one solution
of (\ref{rigg}) for $\vec{\ell}_{-} |_{p^{-}}$ pointing outwards from $\M^{-}$.
\end{lemma}
\begin{proof}
Take two solutions
$\vec{\ell}_{-} |_{p^{-}}$ and $\vec\hatell_{-}|_{p^{-}}$
of (\ref{rigg}). From the second equation it follows that
its difference must be proportional to a normal vector:
$$\vec{\hatell}_{-} |_{p^{-}}= \vec{\ell}_{-} |_{p^{-}} +
A \vec{\nany}^{-}|_{p^{-}}.$$
Inserting this into the first equation in (\ref{rigg}) we obtain
\begin{equation}
0= A (A \nany^{-}_{\mu} \nany^{-\mu} + 2  \nany^{-}_{\mu} \ell_{-}^{\mu}),
\label{eqN}
\end{equation}
which admits two solutions. The solution with $A \neq 0$ gives
an $\vec\hatell_{-}$ satisfying 
$\nany^{-}_{\mu} \hatell^{\mu}_{-} |_{p^{-}} =
- \nany^{-}_{\mu} {\ell}^{\, \mu}_{-} |_{p^{-}}$. 
Thus, $\vec{\ell}$ and $\vec\hatell$ cannot both have the correct orientation.
\finn
\end{proof}

\vspace{3mm} 

The next Lemma shows that, at null points, uniqueness of 
$\vec{\ell}_{-} |_{p^{-}}$ holds {\em irrespective} of orientation. 
\begin{lemma}
\label{uniquenessNull}
With the same notation as in Lemma \ref{uniquenessNonNull} assume now that
$\S^{-}$ is null at $p^{-}$. Then the
solution of the algebraic equations (\ref{rigg}) at $p^-$ is unique,
if it exists.
\end{lemma}
\begin{proof}
As $\nany^{-}_{\mu} \nany^{-\mu}|_p=0$, 
equation (\ref{eqN}) simplifies to $0 = 2 A \nany^{-}_{\mu}
\hatell_{-}^{\mu} |_{p^{-}}$. Transversality of the rigging immediately implies then that
$A=0$. 
\finn
\end{proof}

\vspace{3mm}

This Lemma implies that the orientation of
$\vec{\ell}_{-}$ is {\em fixed} directly by the algebraic
conditions (\ref{rigg}) at null points. This
clearly suggests that there will exist
spacetimes-with-boundaries satisfying all the
preliminary matching conditions which, however, cannot be matched
continuously. 

Before showing this explicitly, we must check that the existence
of an $\vec{\ell}_{-}$ does not depend on the choice of
$\vec{\ell}_{+}$.  Assume that a solution of the preliminary matching
conditions (\ref{prelim}) exists for one choice of rigging $\vec{\ell}_{+}$ and take
any other rigging $\vec\hatell_{\, +}$. To show that a solution also
exists for the second choice, we only need to decompose
$\vec\hatell_{\, +}$ in the basis
$\{ {\vec{e}}_{a}^{\, +}, \vec{\ell}_{+} \}$
and {\it define} $\vec\hatell_{-}$ as the same linear combination of 
$\{ {\vec{e}}_{a}^{\, -}, \vec{\ell}_{-} \}$ (with $\vec{\ell}_{-}$ being
the solution 
for $\vec{\ell}_{+}$ which we assume it exists and which we know it is unique).
All the rigging and orientability conditions for $\vec\hatell_{\pm}$ are
automatically satisfied.
Thus, existence (or non-existence) of a
suitable pair of riggings is reduced to existence of a solution of
(\ref{rigg}) for $\vec{\ell}_{-}$ given
{\it any chosen} rigging $\vec{\ell}_{+}$. 

We can now discuss
examples showing that the preliminary matching conditions are not
sufficient for the existence of a continuous matching.
Let us begin with the simplest possible example, so that
the main obstruction to existence becomes clear.
Let us consider two identical copies of the submanifold-with-boundary
defined by $t\geq x$ in 2-dimensional Minkowski spacetime in Cartesian coordinates $\{t,x\}$.
Let us denote them 
by $(M^+,\eta)$ and $(M^-,\eta)$. Their corresponding boundaries are
obviously  $\S^{\pm}:\{t=x\}$, see  Figure \ref{fig:fig1}. 
Let us now try to match them by identifying
the boundaries in the natural way, i.e. by taking $\Phi$ as the
identity mapping. Without loss of generality, let 
the rigging vector $\vec \ell_+$ be null and point towards
$M^+$. We know by (\ref{rigg}) that $\vec \ell_-$ also has to be null.
Moreover it has to point outwards from $M^-$ (see Figure \ref{fig:fig1}).
However, with the natural identification we have chosen, if the tangent
vector $\vec{e}^{\,+}_{1}$ points in one possible direction, then the
tangent vector $\vec{e}^{\, -}_{1}$ to be identified with $\vec{e}^{\,+}_{1}$ must 
also point in that same direction, see Figure \ref{fig:fig1}. But then it is clear that 
the second equality in (\ref{rigg}) cannot be satisfied, showing that
these two spacetimes cannot be matched across their boundaries by using the
natural identification of points. One might still think that
the problem arises from the choice of identification of
boundaries. This is not the case, however, because generically two
manifolds-with-boundary will have at most {\it one} diffeomorphism between
their boundaries for which the preliminary matching conditions are satisfied.
Hence, in general there is no freedom in choosing another identification
(see Corollary \ref{res:noiso} below).

%Flipping Figure \ref{fig:fig1} over $\S$,
%(see (c) in Fig.\ref{fig:fig1})
%to try to match them, turns out to be unsatisfactory, because

\setlength{\unitlength}{1mm}
\begin{figure*}[th]
\centering
\begin{picture}(100,50)
%\graphpaper(0,0)(150,50)
\includegraphics[width=5cm]{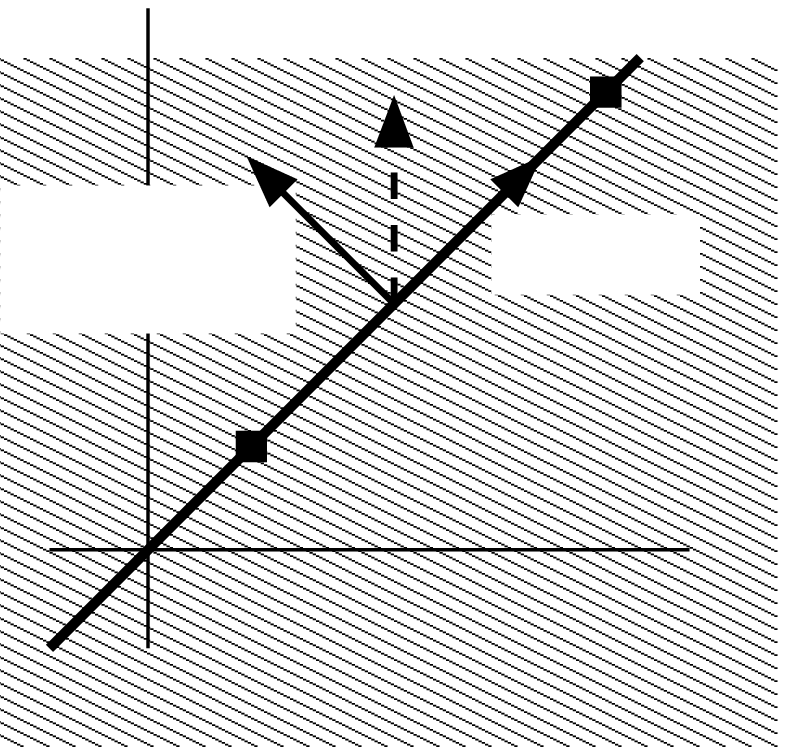}
\includegraphics[width=5cm]{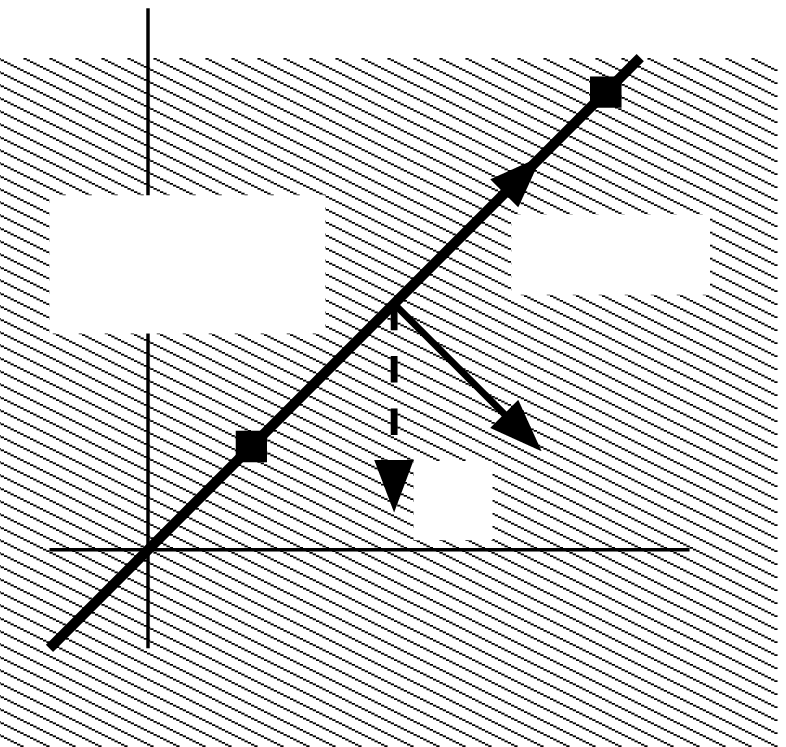}
% \put(-146,45){$t$}
% \put(-95,45){$t$}
% \put(-41,45){$x$}
% \put(-107,8){$x$}
% \put(-55,8){$x$}
% \put(-5,8){$t$}
% \put(-124,36){$\vec n$}
% \put(-73,36){$\vec n$}
% \put(-13,34){$\vec n$}
% \put(-133,36){$\vec \ell_+$}
% \put(-74,15){$\vec \ell_-$}
% \put(-35,31){$\vec \ell_-$}
% \put(-141,19){$a$}
% \put(-90,19){$a$}
% \put(-30,16){$a$}
% \put(-118,43){$b$}
% \put(-67,43){$b$}
% \put(-7,38){$b$}
% \put(-150,30){$(\M^+,\eta)$}
% \put(-95,30){$(\M^-,\eta)$}
% \put(-20,20){$(\M^-,\eta)$}
\put(-95,45){$t$}
\put(-44,45){$t$}
\put(-57,8){$x$}
\put(-6,8){$x$}
\put(-70,30){$\vec \nany^+=\vec e_1^+$}
\put(-18,30){$\vec \nany^-=\vec e_1^-$}
\put(-83,38){$\vec \ell_+$}
\put(-23,14){$\vec \ell_-$}
\put(-90,19){$a$}
\put(-39,19){$a$}
\put(-67,43){$b$}
\put(-16,43){$b$}
\put(-100,30){$(M^+,\eta)$}
\put(-49,30){$(M^-,\eta)$}
\end{picture}
\caption{Two identical copies of the manifold with boundary defined by
the region $t\geq x$ of Minkowski spacetime.
These are the non-shadowed parts of the picture. The natural
identification of boundaries amounts to identifying the two $a$-points,
and the two $b$-points, and so on. The rigging vectors
$\vec\ell^{\pm}$ are chosen to be null and
one pointing towards $M^+$ and the other outwards from $M^-$, as required.
The second vector $\vec{e}^{\, \pm}_{1}$ to complete the bases
$\{\vec\ell^{\pm},\vec{e}^{\, \pm}_{1}\}$ coincides with the
corresponding normal vector $\vec N^{\pm}$. Both $\vec{e}^{\, \pm}_{1}$
must point from $a$ to $b$ (or vice versa) due to the chosen identification.
But then the tangent spaces cannot be properly identified because the
scalar products do not match. A second possible choice of
rigging vectors is represented by the broken-line vectors,
leading to the same problem.}
\label{fig:fig1}
\end{figure*}

This example can be generalised to arbitrary spacetimes. First of all
let us notice that a natural way of building spacetimes-with-boundary 
is picking up an arbitrary spacetime $(\V,\gamma)$ and choosing a hypersurface $\S^{-}$
which divides $\V$ into two regions, which we denote by $\M^{-}_1$ and 
$\M^{-}_2$. Clearly both regions have $\S^{-}$ as their boundary. 
Assume now that we have another spacetime-with-boundary
$(\M^{+},g^+)$ and assume that $\M^{+}$ can be matched with $\M^{-}_1$ (say)
to produce a spacetime with continuous metric $g$. 
The following proposition ensures that $\M^{+}$ {\it cannot} be matched
to $\M^{-}_2$ using the same identification of boundaries if $\S^{-}$ has at least one null
point.
\begin{proposition}
\label{OnlyOneWorks}
Let $(\M^{+},g^{+})$ be a $C^2$ spacetime with boundary $\S^+$
and $\M^{-}_{1}$, $\M^{-}_2$ be two regions of a $C^2$ spacetime $(\V,\gamma)$
satisfying 
$$\M^{-}_{1} \cup \M^{-}_{2} = \V, \hspace{1cm} 
\M^{-}_{1} \cap \M^{-}_{2} = \S^{-}
$$ 
where $\S^{-}$ is
a $C^3$ hypersurface
with at least one null point. 
%Denote by $g^{-}_{1}$ ($g^{-}_{2}$)
%the natural induced metrics on $\M^{-}_1$ ($\M^{-}_2$). 
If there exists a diffeomorphism $\Phi$ between 
$\S^{+}$ and $\S^{-}$ such that 
$(\M^{+},g^{+})$ can be matched continuously to 
$(\M^{-}_1,\gamma|_{\M^{-}_1})$, then
$(\M^{+},g^{+})$ cannot be matched continuously to $(\M^{-}_2,\gamma|_{\M^{-}_2})$ with
the same diffeomorphism $\Phi$.
\end{proposition}
\begin{proof}
Take a point $p^{-} \in \S^{-}_0$ %where $\S^-$ is null
and define $p^{+} = \Phi(p^{-})$ as usual. By Lemma \ref{uniquenessNull},
for any transverse vector $\vec{\ell}^{+} |_{p^+}$ pointing towards $\M^{+}$
there exists exactly one transverse vector $\vec{\ell}^{-} |_{p^{-}}$
satisfying the rigging conditions (\ref{rigg}). Moreover, we know that
$\vec{\ell}^{-} |_{p^{-}}$ points outwards from $\M^{-}_1$ because
$(\M^{+},g^{+})$ can be matched continuously to
$(\M^{-}_1,\gamma|_{\M^{-}_1})$. Thus, there exists no rigging solving
(\ref{rigg}) pointing towards $\M^{-}_{1}$. Since $\M^{-}_1$ and
$\M^{-}_2$ can be visualized inside the total spacetime $\V$, it follows that
there is no rigging solving (\ref{rigg}) pointing outwards from $\M^{-}_2$.
%Thus, $(\M^{-}_2,g^{-}_2)$ cannot be matched with $(\M^{+},g^{+})$ to
%produce a spacetime with a continuous metric.
\finn
\end{proof}

\vspace{3mm}

For any hypersurface $\S$ of arbitrary causal character, with
first fundamental form $h$, a diffeomorphism $\Psi :
\S \rightarrow \S$ is called an {\it isometry} if 
$\Psi^{\star} (h) = h$.  The following corollary follows immediately
from Proposition \ref{OnlyOneWorks}, taking into account
that if $(\M^{+},g^{+})$ could still
be matched to $(\M^{-}_2,\gamma|_{\M^{-}_2})$ through a different diffeomorphism
$\Phi'$, then $\Phi^{-1}\circ \Phi'$ would constitute an isometry
of $\S^-$ different from the identity.

\begin{corollary}
\label{res:noiso}
With the same assumptions as in Proposition \ref{OnlyOneWorks}, let
$h^{-}$ be the first fundamental form of $\S^{-} \in \V$. If
$(\S^{-},h^{-})$ admits no isometries apart from the identity,
then $(\M^{+},g^{+})$
cannot be matched to $(\M^{-}_2,\gamma|_{\M^{-}_2})$.
\end{corollary}

\subsection{Gluing and the $Z_2$-mirror symmetry}
Proposition \ref{OnlyOneWorks} has another interesting corollary 
applicable to the case of joining two identical copies of the same 
$C^2$ spacetime with boundary: the so-called $Z_2$-mirror symmetry branes. 
If the boundary has at least one null
point, and if the spacetime is a subregion of a larger $C^2$
spacetime without
boundary, impossibility of matching would follow immediately. It is
likely that such a $C^2$ extension always exists. However, instead of
attempting a proof of this fact, let us show directly that, in any
case, the two copies cannot be matched continuously, as announced in \cite{letter}.

\begin{corollary} 
\label{res:2} 
It is impossible to join two identical copies of a spacetime with boundary $\S$ such that $\S_0\neq \emptyset$ (in particular with a signature-changing boundary $\S$), identifying naturally the points on
$\S$, to produce a bulk with continuous metric.
\end{corollary} 
\begin{proof}
Let us call $(\M^{\pm},g^{\pm})$ the two identical copies and
$\S^{\pm}$ their corresponding boundaries. Let $\chi$ be the
natural identification of $\M^{+}$ and $\M^{-}$.  Take a point $p$
where the boundary is null and any rigging vector $\vec{\ell}_{+} |_p$
pointing towards $\M^{+}$. Clearly $\chi |_{\S_+}$ is the
diffeomorphism we are using to identify the boundaries.
By Lemma \ref{uniquenessNull} there is a unique
solution $\vec{\ell}_{-} |_{\chi(p)}$ of the rigging equations (\ref{rigg}),
and that must be obviously
$\vec{\ell}_{-} |_{\chi(p)} \equiv \d\chi ( \vec{\ell}_{+} |_p )$ which
is just a copy of the original rigging. But since $\vec{\ell}_{+} |_p$
points towards $\M^{+}$ it follows that $\vec{\ell}_{-} |_{\chi(p)}$
points towards $\M^{-}$ and the proposed matching is impossible.
\finn
\end{proof}

\vspace{3mm} 

This corollary shows that the usual $Z_2$ mirror symmetry extensively used
in the brane scenario is {\it incompatible} with signature-changing
branes, with null branes, and in general with branes having a non-empty $\S_0$. Therefore constructing such branes requires more
sophisticated methods. For instance, we can try to join
two different regions of the
same spacetime or two different spacetimes. In general such constructions
are more involved than in the mirror symmetric case 
because the preliminary matching conditions are not
automatically satisfied and more equations
need to be solved. This does not mean, however, that such constructions
are impossible. Explicit examples were in fact given in \cite{letter}.
We shall go back to these and other examples in Sections \ref{static} and \ref{AdSect}. 

In this section we have seen that in order to construct
spacetimes with signature changing branes one must be careful with the
existence of suitable riggings at points where the matching
hypersurface is null. The results we have presented obviously
hold for usual matching conditions in General Relativity,
but they also hold in any other geometrical theory.
When trying to join spacetimes
involving matching hypersurfaces with null points (in particular if
the matching hypersurface is null everywhere), the equality of the
first fundamental forms {\it is not sufficient} to ensure the
existence of a matched spacetime with continuous metric. Existence of
suitable riggings must always be checked in those cases.

Having discussed the construction of branes by the
method of gluing and its consequences for the signature-changing case, 
we need to analyze the equations relating the
jump in the metric derivatives with the singular part of the Einstein tensor
on the bulk. Again, the standard Israel
conditions do not apply in the signature changing case and need to be generalised. We
discuss the results in the next section.

\section{Generalised Israel formula}
\label{genIsrael}

Under the assumptions of Theorem \ref{theorem1} we have an orientable $C^3$
bulk $\M$ with a continuous, piecewise $C^2$, metric $g$.
We choose an orientation 
on $\M$ and denote by $\bm{\eta}$ its canonical volume $n$-form. This allows us to define the
Riemann, Ricci and Einstein tensors in a distributional way. Since the definitions of the Riemann and
Einstein tensors contain
second derivatives of the metric and this is not, in general, $C^1$ across
$\S$, one expects that delta-type distributions 
with support on $\S$ will arise. Indeed, it can be shown \cite{ClarkeDray87,MarsSenovilla93}
that the Einstein tensor of $g$, viewed as a tensor distribution on $\M$ and denoted
by $\und{G}_{\mu\nu}$, takes the form
\begin{equation}
\und{G}_{\mu\nu}
=\otheta^{+} G^+_{\mu\nu} + 
\otheta^{-}  G^-_{\mu\nu}
+\delta\, \tt_{\mu\nu},
\label{ddEin}
\end{equation}
where $G^{\pm}_{\mu\nu}$ are the Einstein tensors of $(\M^{\pm},g^{\pm})$
and $\tt_{\mu\nu}$, which is defined only on $\S$, 
is called {\it the singular part
of the Einstein tensor distribution}.  The distributions 
$\otheta^{\pm}$ and $\delta$
are defined as follows: for any test function $Y$
(i.e. a $C^3$ function with compact support on $\M$ ---note that $\M$ is
only $C^3$ so it makes no sense to assume higher differentiability for $Y$---)
\[
\left < \otheta^{\pm}, Y \right > \equiv
\int_{\M^{\pm}} Y \bm{\eta}\, .
\]
Regarding $\delta$, we first define
a one-form distribution $\bm\delta \equiv \nabla \otheta^+=-\nabla\otheta^-$, see \cite{MarsSenovilla93}. Explicitly,
$\bm\delta$ acts on any test vector field $\vec Y$ ($C^2$
vector field with compact support) as
\[
  \left< \bm \delta,\vec Y\right>=\int_{\S} Y^\mu d\sigma_\mu,
\]
where $d\sigma_\mu$ is 
%a one-form on $\M$ taking values on $(n-1)$-forms on $\S$ explicitly 
defined as
$$
d\sigma_{\mu} = 
\eta_{\mu \alpha_1\ldots \alpha_{n-1}}e^{\alpha_1}_1\ldots
e^{\alpha_{n-1}}_{n-1} d\xi^1 \wedge \cdots \wedge d \xi^{n-1}$$
where
$\vec{e}_a = \vec{e}_a^{\, +} = \vec{e}_a^{\, -}$,
%are defined as in (\ref{e's}) 
and $\{ \vec{\ell},\vec{e}_1, \cdots \vec{e}_{n-1} \}$ is
a positively-oriented basis, that is $\ell^\mu d\sigma_\mu >0$ 
(recall that $\vec{\ell}= \vec{\ell}_{+} = \vec{\ell}_{-}$ after
identification).
%For reasons that will become clear later,

It is convenient here to
choose the normal
\[
\bm \normal\equiv \frac{1}{\nany_\alpha \ell^\alpha}\bm \nany,
\]
which does not depend on the choice of $\bm\nany$,
but does indeed on the choice of rigging vector $\vec\ell$:
given the rigging, its intrinsic characterisation is
$\bm\normal\propto \bm\nany$ and
$\normal_\alpha\ell^\alpha=1$.
The importance of this normal is that the identification
of the tangent vectors and the riggings at both sides
induces the identification $\bm \normal=\bm \normal^+=\bm \normal^-$.

Denoting by $d\sigma$ a volume element on $\S$
defined by 
$$d\sigma_\mu= \normal_\mu d\sigma \hspace{2mm} \Longleftrightarrow \hspace{2mm}
d\sigma=\ell^\alpha d\sigma_\alpha
$$
%$\sign(n_\alpha \ell^\alpha) \normal_\mu d\sigma$,
the distribution $\delta$ is defined by 
\[
\left < \delta, Y \right >\equiv \int_{\S} Y d\sigma \enspace,
\]
where $Y$ is any test function. $\delta$ obviously depends on
the choice of rigging
via the normal form $\bm \normal$.
From the identity
$$
\bm \delta= \bm \normal \, \delta
$$
and the fact that $\bm{\delta}$ is intrinsically defined, it follows
that a rescaling $\vec\ell'=A^{-1}\vec\ell$, so that
$\bm \normal'=A \bm \normal$, induces the transformation
\begin{equation}
\delta'=A^{-1} \delta, ~~~~\Longrightarrow ~~~~
\tt'_{\mu\nu}=A\;\tt_{\mu\nu},
\label{eq:rescal}
\end{equation}
after using (\ref{ddEin}).
Observe that both $\bm{\delta}$ and $\delta$ have support on $\S$.

We still need to specify the explicit form of $\tt_{\mu\nu}$
in expression (\ref{ddEin}).
By construction, the metric $g$ has continuous 
tangential derivatives at $\S$. Therefore, this singular
part of the Einstein tensor distribution
will be related to the
discontinuity of the transversal derivatives of the metric across 
$\S$. In the cases where $\S$ is timelike (or spacelike) everywhere, the normal vector
$\vec{\nany}$
is transversal to $\S$, and therefore we can choose the rigging
to be proportional to $\vec{\nany}$ and unit.
This implies that $\bm{\normal}$
is also unit and in fact $\ell^{\alpha} = \sign(\bm{n}, \bm{n}) n^{\alpha}$.
Thus, the second fundamental forms inherited by $\S$ from both sides, which can be
promoted to spacetime tensors by means of the definition
\[
K^\pm_{\mu\nu}\equiv \proj_\mu^{\alpha} \proj_{\nu}^{\beta}
\nabla^\pm_\alpha \normal_\beta,
\]
where $\nabla^{\pm}$ are the Levi-Civita connections of $g^{\pm}$
and 
$$\proj^{\alpha}_{\mu} =
\delta^{\alpha}_{\mu} - \sign(\bm{n}, \bm{n}) \normal^{\alpha} \normal_{\mu}
$$ 
is the projector orthogonal to $\S$,
encode properly the
jumps of the transversal derivatives of the metric. 
It is not surprising, therefore, that 
$\tt_{\mu\nu}$ can be written down in terms
of the jumps of the second fundamental forms in the non-null case. 
This is the content
of the so-called Israel formula \cite{Israel66} which reads,
taking $\sign (\bm{\normal}, \bm{\normal})
\vec{\normal}$ pointing towards $\M^{+}$,
\begin{equation}
%\sign(\bm{n}, \bm{n})~
\tt_{\mu\nu}=  -\left [ K_{\mu\nu} \right ] + \proj_{\mu\nu} \left [K \right ].
\label{Israel}
\end{equation}
%where here $h_{\mu\nu} =g_{\mu\alpha} P^\alpha_\nu$
%reads
% $g_{\mu\nu}-(\bm{n}, \bm{n})^{-1} n_{\mu} n_{\nu}$
Here and in the rest of the paper, the
``discontinuity'' $[f]$ of any object which has well-defined limits at both sides of $\S$ is defined as
\[
\left[ f \right ] (p) \equiv
\mathop{\lim} \limits_{x \mathop \to \limits_{\M^{+}}p} f^{+}(x) -
\mathop{\lim} \limits_{x \mathop \to \limits_{\M^{-}}p} f^{-}(x) 
\hspace{1cm} \ \forall p \in \S.
\]

In the signature-changing case, and in general whenever $\S_0\neq \emptyset$, the normal
vector is no longer transverse to the hypersurface everywhere.
Thus, the second fundamental forms of $\S$ are no longer suitable to
measure the jumps in the transversal derivatives of the metric.
This makes clear that the Israel formula (\ref{Israel}) must be modified in
these cases. Taking into account
that $\vec{\ell}$ is transverse to the
hypersurface, it is natural to substitute the second fundamental
forms by the new objects \cite{MarsSenovilla93} 
\footnote{It must be
remarked that, in purity, the second fundamental form of a hypersurface
is a tensor field defined only on the hypersurface. Thus, the rigorously
defined object is in fact
$K_{ab}=-g\left(\vec n , \nabla_{\vec e_a}\vec e_b\right)$, which is symmetric.
One can however use any unit extension of $\bm{n}$ outside $\S$ to define
$K_{\mu\nu}$ and then, in fact, $K_{ab}=K_{\mu\nu}e^{\mu}_a e^{\nu}_b$.
Similarly, the rigorously defined object using the rigging is
$\hh_{ab}=-g\left(\vec \ell , \nabla_{\vec e_a}\vec e_b\right)$, which
in this is case is not necessarily symmetric. Extensions of $\bm{n}$ and $\vec\ell$ outside $\S$ keeping $n_{\mu}\ell^{\mu}=1$ permit then to define $\hh_{\mu\nu}$ and, again, $\hh_{ab}=\hh_{\mu\nu}e^{\mu}_a e^{\nu}_b$.}
\begin{equation}
\left. \hh^{\pm}_{\mu\nu}\equiv \Pi^{\alpha}{}_{\mu}\Pi^{\beta}{}_{\nu}
\nabla^{\pm}_\alpha {\ell}_\beta \right |_{\S},
\label{hh}
\end{equation}
where now the generalized projector $\Pi^{\alpha}{}_{\mu}$ reads
$$
\Pi^{\alpha}{}_{\mu}=\delta^{\alpha}_{\mu}-n_{\mu}\ell^{\alpha}\, .
$$
Observe that $\Pi_{\mu\nu}$ is no longer symmetric, and that $\ell^\mu\Pi^{\alpha}{}_{\mu}=0$, hence
$$
\ell^\mu\hh^{\pm}_{\mu\nu}=0, \,\,\,\,\, \ell^\nu\hh^{\pm}_{\mu\nu}=0.
$$ 

We can now write down the expression of
$\tt_{\mu\nu}$ in terms of jumps of these objects \cite{MarsSenovilla93} 
%for the derivation, but note
%the above discussion and the footnote)
%that here we are not imposing $n_\alpha\ell^\alpha=1$)
\begin{eqnarray}
\tt_{\mu\nu}& = &%\frac{1}{|\ell^\lambda \normal_\lambda|}
\normal^\alpha[\hh_{\alpha\mu}]\normal_\nu+
\normal^\alpha[\hh_{\alpha\nu}]\normal_\mu\nonumber\\
&&-
\normal^\alpha \normal_\alpha[\hh_{\mu\nu}]-
\normal_\mu \normal_\nu[\hh^\alpha{}_\alpha]\nonumber\\
&&-\g_{\mu\nu}|{}_{{}_{\S}}
\left(\normal^\alpha \normal^\beta[\hh_{\alpha\beta}]-
\normal^\alpha \normal_\alpha[\hh^\beta{}_\beta]\right).
\label{eq:tt}
\end{eqnarray}
This is the generalization of Israel's formula (\ref{Israel})
to arbitrary hypersurfaces. The transformation (\ref{eq:rescal}) under
change 
of rigging can be directly checked in this expression,
taking into account that $[\hh'_{\alpha\beta}]=A^{-1}[\hh_{\alpha\beta}]$.
Note that \cite{MarsSenovilla93}: 
\begin{itemize}
\item $\hh^\pm_{\mu\nu}$ are not symmetric, but $[\hh_{\mu\nu}]$ is symmetric;
\item $[\hh_{\mu\nu}]$ does not depend on a change of rigging
$\vec \ell'=\vec \ell+f^a \vec e_a$ for any functions $f^a$ defined on
$\S$. Neither $\bm \normal$ does, and therefore $\tt_{\mu\nu}$ does not
depend on a change of rigging of that kind.
Thence, the only transformation of
%dependency of
$\tt_{\mu\nu}$ under a change of rigging
is through the rescaling (\ref{eq:rescal}) discussed above.
%\item Eq.(\ref{eq:tt}) remains valid for any choice of rigging
%used to define  $\hh^\pm_{\mu\nu}$.
\end{itemize}
These are of course necessary consistency
properties of the final expression (\ref{eq:tt}).
It should also be remarked that, for timelike branes, this expression
reduces to the usual Israel formula by taking $\vec{\ell} = \sign
(\bm{\normal},\bm{\normal}) \vec{\normal}$ as before.
Note that then
$\normal^\alpha \hh^\pm_{\alpha\beta}=\ell^\alpha \hh^\pm_{\alpha\beta}=0$
and $\hh^\pm_{\alpha\beta} = \sign (\bm{n}, \bm{n}) K^\pm_{\alpha\beta}$.
%(this is most easily seen by taking the rigging to be precisely
%$\vec{n}$ in the definition of $\hh_{\mu\nu}$, taking into account that then
%the two normals are identified, and that $P^\alpha_\beta= w^a{}_\beta e_a{}^\alpha$).

The generalised expression (\ref{eq:tt}) satisfies
$$
\normal^{\mu} \tt_{\mu\nu} \equiv 0
$$
as one can immediately check.
%still defines a %n energy-momentum
%tensor {\it on the brane}
%in the sense that $\normal^{\mu} \tt_{\mu\nu} \equiv 0$ . 
Thus, at points where $\S$ is not null
only the {\it tangential} components $\tt_{ab}=e_a{}^\alpha e_b{}^\beta\tt_{\alpha\beta}$ 
are present, and they contain \emph{all} the information carried by $\tt_{\mu\nu}$.
%These are $\tt_{ab}=e_a{}^\alpha e_b{}^\beta\tt_{\alpha\beta}$.
For a $\S$ with a non-empty $\S_0$,  though, one should bear in mind that the
normal vector $\vec{\normal}$ is tangent to $\S$ at the null phase $\S_0$, so that the geometrical interpretation of this vanishing contraction is not so straightforward.

\section{Field Equations: the energy-momentum tensor on the brane}
\label{fieldeqs}
We are now in a position where the Einstein equations on the bulk can be 
discussed. Due to the structure of the Einstein tensor distribution of the bulk
(\ref{ddEin}), the corresponding energy-momentum tensor
on the bulk $\ttm_{\mu\nu}$ will also be a tensor distribution and will
consist of three parts:
the tensor fields $T^{\pm}_{\mu\nu}$ defined on each region $\M^{\pm}$,
at each side of $\S$, plus a singular part with support on $\S$
proportional to $\delta$, 
\[
\ttm_{\mu\nu}=\otheta^{+} T^+_{\mu\nu}+
\otheta^{-} T^-_{\mu\nu}+
\delta\, \tau_{\mu\nu}.
\]
Notice that, again,
$\ttt_{\mu\nu}$ does not have intrinsic meaning because $\vec{\normal}$
cannot be canonically normalized on a signature-changing brane. Only the
product $\delta\, \ttt_{\mu\nu}$ is well-defined. For the individual term $\ttt_{\mu\nu}$ to
become meaningful, a volume element must be fixed once and for all on $\S$. 
Equivalently, one must choose a given rigging, which in turn determines a unique normal one-form.
Despite these issues, we will refer to $\ttt_{\mu\nu}$ as the {\em energy-momentum tensor on the brane $\S$}.

Keeping this in mind, the Einstein equations on
the bulk are given by
\begin{equation}
\und{G}_{\mu\nu}+  \und \Lambda_{\mu\nu}=\kappa^2_n\; \ttm_{\mu\nu},
\label{EFE}
\end{equation}
where $\kappa_n$ is the $n$-dimensional gravity coupling
constant and we have set
\[
\und{\Lambda}_{\mu\nu}
=\otheta^{+} \Lambda_n^+ g^+_{\mu\nu}+
\otheta^{-}  \Lambda_n^- g^-_{\mu\nu},
%+\delta\cdot \tension.
\label{eq:5d_lambda}
\]
where $\Lambda^{\pm}_n$ are the cosmological constants corresponding to $\M^{\pm}$. Observe that we are allowing for {\em different} values of the cosmological constant at each side of the brane $\S$.

The Einstein equations (\ref{EFE}) decompose then as
\[
G^{\pm}_{\mu\nu} + \Lambda_n^\pm g^{\pm}_{\mu\nu}= \kappa_n^2 T^{\pm}_{\mu\nu}
\]
on each of $\M^{\pm}$ plus
\begin{equation}
\tt_{\mu\nu} = \kappa_n^2 \ttt_{\mu\nu}
\label{eq:brane_equation}
\end{equation}
at points on $\S$.
Let us insist once more that this last equation is intrinsic
only when multiplied by the distribution $\delta$. However, one can still write (\ref{eq:brane_equation}) as it stands because both sides of the equation are affected by exactly the same normalization freedom. Furthermore, note that 
(\ref{eq:brane_equation}) together with (\ref{eq:tt})
constitute the generalisation of the Israel equations
to general hypersurfaces
in terms of the energy-momentum of the (hyper)surface layer.

The geometrical property $\normal^\mu \tt_{\mu\nu}=0$
implies then that 
\begin{equation}
\normal^\mu \ttt_{\mu\nu}=0.\label{nova}
\end{equation}
As discussed above, at points where $\S$ is not null,
in particular on its Lorentzian part $\S_L$,
equations (\ref{eq:brane_equation}) are equivalent to
the $n(n-1)/2$ projected equations
\begin{equation}
\label{eq:eqs_on_brane}
\tt_{ab} = \kappa_n^2 \ttt_{ab},
\end{equation}
which are defined on the brane, where as usual
$$
\ttt_{ab} =e_a{}^\mu e_b{}^\nu\ttt_{\mu\nu} .
$$
Nevertheless, for general branes the $n(n-1)/2$ independent relations contained in (\ref{eq:brane_equation}) are not so simply interpreted, and in fact the meaning of (\ref{nova}) on the null phase $\S_0$ and the signature-changing set $S\subset \S_0$ is that any {\em tangential} component of $\ttt_{\mu\nu}$ along the unique null degeneration direction must vanish.

% \begin{proposition}
% Let $\S$ be a hypersurface on a spacetime $\M$
% which changes its causal character on a $(n-2)$-dimensional surface $S$.
% Then 
% \begin{eqnarray*}
% \mathop{\lim} \limits_{ x   \mathop \to S}
% \frac{K_{ab} n^a n^b}{\left (\bm{n},\bm{n} \right)} \rightarrow \infty,
% \end{eqnarray*}
% where $K_{ab}$ is the second fundamental form with respect to $\bm{n}$ and
% $n^a$ is defined by (\ref{decomn}).
% \end{proposition}

% {\it Proof:} 

% \begin{eqnarray*}
% \partial_a \left (\bm{n},\bm{n} \right )  =
% - 2 \varphi_a + 2 K_{ab} n^b,
% \end{eqnarray*}
% where
% \begin{eqnarray*}
% \varphi_a =  \frac{1}{\ell^{\alpha} n_{\alpha} }
% \ell^{\nu} e_a^{\mu} \nabla_{\mu} n_{\nu}
% \end{eqnarray*}

It is customary to decompose the total
energy-momentum tensor on the Lorentzian part $\S_L$ of the brane into two parts \cite{roy_review}:
the {\em brane tension} which takes the form $-\tension h_{ab}$ of a cosmological constant term given by some effective theory defining the brane,
and the energy-momentum tensor $\ttt^{m}_{ab}$ of the particles and fields confined
to the brane. Following the same idea, sometimes
we will %make let us therefore
consider a similar decomposition {\em all over} $\S$
\begin{equation}
\label{eq:tensionandmatter}
\ttt_{ab}=-\tension h_{ab}+\ttt^{m}_{ab}\, .
\end{equation}
It must be remarked then
that, at points in the signature-changing set $S$, $\ttt^{m}_{ab}$
does not contain the full information of the energy-momentum tensor of fields ``confined'' on the brane.

Regarding specific energy-momentum tensors on the brane,
much attention has been focused to the case where the
total energy-momentum tensor on the brane is
of ``cosmological constant type'', probably for simplicity.
From relation (\ref{eq:tensionandmatter}) it follows that this case corresponds
to a brane with non-vanishing tension but no matter content, so that
$\ttt^{m}_{ab}=0$.
%$\alpha=-\tension$ and
%, this is, when
%$\tau_{ab} =\alpha h_{ab}$.
In the final part of this section
we will show that for signature-changing branes the
energy-momentum tensor {\it cannot} be of this type near $S$. 
We do this in two steps: for so-called umbilical branes, and
in the general case.

\subsection{Umbilical branes}
Recall that a hypersurface is called {\it umbilical} whenever
its second fundamental form is proportional to the first fundamental form:
$K_{ab}\propto h_{ab}$. In the constant-signature case, the most simple way of
obtaining $\ttt_{ab}=\alpha h_{ab}$ for some scalar field $\alpha$ consists on
gluing two umbilical hypersurfaces $\S^\pm$. This follows immediately from the
standard Israel formula (\ref{Israel}). As a matter of fact, this procedure is
{\em exclusive} of constant-signature  branes, because signature-changing branes
cannot be umbilical, as we show next.

To that end, let us decompose 
the normal vector $\vec{\normal}$ in the basis
$\{\vec{\ell}, \vec{e}_a \}$. Since 
the contraction of 
$\vec{\normal} - (\bm{\normal}, \bm{\normal} ) \vec{\ell}$
with $\bm{\normal}$ vanishes, it follows that this vector
must be a linear
combination of the tangent vectors $\vec{e}_a$. Denoting the
coefficients by $\normal^a$ we have
\begin{equation}
\vec\normal = (\bm{\normal}, \bm{\normal} )\vec{\ell} + \normal^a \vec{e}_a.
\label{decomn}
\end{equation}

Recall also that the second fundamental form
defined as an object in $\S$ reads
$K_{ab}= e_a{}^\mu e_b{}^\nu K_{\mu\nu}=e_a{}^\mu e_b{}^\nu\nabla_\mu \normal_\nu$.
Then, we have the following important result, which was advanced in \cite{letter}. 
\begin{proposition}
\label{umbil}
A $C^3$ umbilical hypersurface 
of a $C^2$ spacetime must have constant signature.
\end{proposition}

\begin{proof}
Let $\S$ be a $C^3$ hypersurface and $\bm \normal$
a $C^2$ normal one-form. Multiplying $\vec{n}$ in (\ref{decomn})
by $\vec{e}_b$ and using $h_{ab}={e_a}^\mu {e_b}^\nu g_{\mu\nu}|_\S$,
it follows
\[
\normal^a h_{ab}=-(\bm \normal,\bm \normal)\ell_\beta{e_b}^\beta.
\]
Defining \cite{MarsSenovilla93}
\[
\varphi_a \equiv  -%\frac{1}{\ell^{\alpha} \nany_{\alpha} }
\ell^{\nu} {e_a}^{\mu} \nabla_{\mu} \normal_{\nu},
\]
it is straightforward to obtain ($\partial_a={e_a}^\mu\partial_{\mu}=\partial/\partial\xi^a$) %\cite{MarsSenovilla93}
\begin{equation}
\partial_a \left (\bm{\normal},\bm{\normal} \right )  =
- 2 (\bm \normal,\bm \normal) \varphi_a + 2 K_{ab} \normal^b.
\label{eq:coi}
\end{equation}
Let us assume now that $\S$ is umbilical, i.e.
\[
%\begin{equation}
%  \label{eq:umbilical}
  K_{ab}=F h_{ab},
%\end{equation}
\]
for some function $F$ on $\S$. $F$
is at least $C^1$, because the second fundamental form
is $C^1$ and $h$ is $C^2$.  
Equation \eqref{eq:coi} becomes
\begin{equation}
  \partial_a \left (\bm{\normal},\bm{\normal} \right )  =
-2\left(\varphi _a + F 
%\frac{F}{\ell^\alpha \nany_\alpha}
\ell_\beta\, {e_a}^\beta\right)(\bm \normal,\bm \normal),
\label{eq:fiu}
\end{equation}
which can be viewed as a differential
equation for $(\bm{\normal},\bm{\normal})$.
Uniqueness of the
solution follows because the first factor on the right hand
side is at least $C^1$ (notice that $\varphi_a$ is $C^1$ from its definition).
Thus, if $(\bm{\normal},\bm{\normal})$ vanishes somewhere, then
it mush vanish everywhere on $\S$. This proves the claim. 
\fin
\end{proof}

\vspace{2mm}

Observe that the door for umbilical hypersurfaces
which are null everywhere is still open.
In this case both the second and first fundamental forms are degenerate and share the null degeneration direction. Thus, one can also try to glue two spacetimes across umbilical null branes.

\subsection{The brane tension}
Let us finally address the question of whether 
there can be general branes with 
%no matter fields but 
only brane tension. Proposition \ref{umbil} is a preliminary
no-go result along that direction.
Nevertheless, in principle one could still try
to obtain $\ttt_{ab}=\alpha h_{ab}$ by gluing two non-umbilical
hypersurfaces. The following result, already announced in \cite{letter}, proves
that such a brane cannot undergo a change of signature unless $\alpha$ vanishes somewhere on $\S$.

\begin{theorem}
\label{res:lambdaterm}
Let $\S$ be a brane constructed under the assumptions of
Theorem \ref{theorem1}. If $\tt_{ab}=\beta h_{ab}$
for a function $\beta$ which is non-zero everywhere on $\S$,
then $\S$ cannot change its causal character.
\end{theorem}
%\mnote{\marc Proof shortened, expression (22) in old version dropped}
\proof
Projecting (\ref{eq:tt}) onto $\S$
with $e_a{}^\mu e_b{}^\nu$
and using $\tt_{ab}=\beta h_{ab}$,
we get
\begin{equation}
\beta h_{ab}=-(\bm\normal,\bm\normal) [\hh_{ab}]
-h_{ab}\left(\normal^\alpha \normal^\beta[\hh_{\alpha\beta}]
-(\bm\normal,\bm\normal)[\hh^\alpha{}_\alpha]\right). \label{eq:Hproph}
\end{equation}
%which implies
%\begin{equation}
%[\hh_{ab}]=\mathcal{F} h_{ab}
%\label{eq:Hproph}
%\end{equation}
%for some function
%$\mathcal{F}$ on $\S$, which must be regular, as $[\hh_{ab}]$ and $h_{ab}$ are.
Expression (\ref{decomn}) and $\ell^{\alpha}\hh^{\pm}_{\alpha\beta}=0$ 
implies $\normal^\alpha \normal^\beta[\hh_{\alpha\beta}]=
\normal^a\normal^b[\hh_{ab}]$. Using also $n^a n^b h_{ab} = 
(\bm\normal,\bm\normal) ( (\bm\normal,\bm\normal) \ell^{\alpha} \ell_{\alpha} -1 )$
which follows by squaring $(\bm\normal,\bm\normal) \ell^{\alpha}$ in (\ref{decomn}),
the contraction of (\ref{eq:Hproph}) with $n^a n^b$ gives
\begin{eqnarray}
(\bm\normal,\bm\normal)\left\{\left(\beta -(\bm\normal,\bm\normal) [\hh^\alpha{}_\alpha]\right)\left\{(\bm\normal,\bm\normal)\ell_{\mu}\ell^{\mu}-1\right\}
\frac{}{}\right.\nonumber\\
\left.\frac{}{}+
\normal^\alpha \normal^\beta[\hh_{\alpha\beta}] (\bm\normal,\bm\normal)\ell_{\mu}\ell^{\mu} \right\}=0.
\label{eq:lambda1}
\end{eqnarray}
Thus, the expression between braces must vanish on $\S_L\cup\S_E$ which readily implies
\[
%\begin{equation}
\lim_{p\rightarrow S} \beta=0.
%-(\bm\normal,\bm\normal)\left\{
%\mathcal{F}(\bm\normal,\bm\normal)\ell_{\mu}\ell^{\mu}
%-[\hh^\alpha{}_\alpha]\right\},
%\end{equation}
\]
Since $\beta$ is at least $C^1$, hence continuous, we have $\beta |_{S} =0$
and the result follows.\fin

%, while from (\ref{eq:Hproph}) we get on $\S_0$
%\begin{equation}
%\beta=
%-\normal^\alpha \normal^\beta[\hh_{\alpha\beta}] \hspace{3mm} \mbox{on} \,\, \S_0. \label{eq:betah}
%\end{equation}
%Using (\ref{eq:Hproph}), (\ref{decomn}) and the fact that $\ell^{\alpha}\hh^{\pm}_{\alpha\beta}=0$, it
%is now straighforward to compute the second term on the right:
%\begin{eqnarray*}
%\normal^\alpha \normal^\beta[\hh_{\alpha\beta}]&=&
%\normal^a\normal^b[\hh_{ab}]= \mathcal{F} \normal^a \normal^b h_{ab}\\
%&=& (\bm\normal,\bm\normal)\mathcal{F}
%\left\{(\bm\normal,\bm\normal)\ell_{\mu}\ell^{\mu}-1\right\}.
% \end{eqnarray*}
%In summary, we have expressions for $\beta$ everywhere on $\S$. From the vanishing of the embraced expression in (\ref{eq:lambda1}) we immediately deduce that, as we approach the signature changing set $S$ from either $\S_L$ or $\S_E$, we must have

Evaluating (\ref{eq:Hproph}) on $S$ and using that $\beta$ vanishes there, we obtain
$$
\normal^\alpha \normal^\beta[\hh_{\alpha\beta}]|_S=
\normal^a\normal^b[\hh_{ab}]|_S = 0 .
$$
In addition to this result, let us note that the identity (see \cite{MarsSenovilla93} for a proof)
$$
[K_{ab}]=(\bm\normal,\bm\normal) [\hh_{ab}]
$$ 
clearly implies that $[K_{ab}]|_{\S_0}=0$ on the null phase $\S_0$; so, if we demand $[K_{ab}]=F h_{ab}$ on $\S$, then $F$ must vanish at the null phase $\S_0$ too.

An important corollary follows from Theorem \ref{res:lambdaterm}
\begin{corollary}
For any choice of normalization,
the condition $\ttt_{ab}=-\Lambda h_{ab}$ for a constant brane tension
$\Lambda\neq 0$ is incompatible with a change of signature on $\S$.
\end{corollary}
A physical interpretation of this result is that a change of signature on the brane requires
that some matter fields become excited, or equivalently
that a signature change cannot occur just spontaneously. Let us remark that the possibility of having 
$\ttt_{ab}=\alpha h_{ab}$ for some {\em function} $\alpha$ has not been ruled out, but this function must necessarily vanish at the signature changing set $S$.
%Let us show that this type of energy-momentum tensors on the brane
%are not supported near the signature changing set.
%ELEVEM AIXO A COROLARI AL FINAL DE LA SECCIO?

% \subsection{The field equations on a signature-changing brane}
% Here we generalise the field equations on the brane as were
% expressed by ... (SMS) in \cite{SMS}
% .
% For that, we relate the Einstein tensor computed from
% the inherited metric on the brane $h_{ab}$ using the Gauss
% and Codazzi equations generalised to a general
% character changing hypersurface \cite{MarsSenovilla93}.
% First important thing to note is that the Ricci scalar,
% and hence the Einstein tensor, cannot be defined as usual
% at points of change of signature, where $\ff=0$.
% This is simply because the induced metric $h_{ab}$ has no inverse
% there. To deal with this problem we will make use of the
% rigged metric $g^{ab}$, which is defined as
% \[
% g^{ab}\equiv g^{\mu\nu} w^a{}_\mu w^b{}_\nu. 
% \]
% This tensor defined all over $\S$ clearly depends on the choice of
% rigging, and can be proven to be degenerate only where $\vec \ell$
% is null. At points where $\ff=0$ and when the rigging is chosen
% in the direction of $\vec n$, then $g^{ab}$ corresponds
% to the inverse of $h_{ab}$...
% The relevant equations read

% JA CONTINUARE AQUI, ENCARA QUE NO TINC MOLT CLAR SI VAL LA PENA...

\section{General branes in static and
spherically, plane, or hyperbolically symmetric bulks}
\label{static}
Our aim now is to provide examples of sufficient generality for the construction described 
in the previous sections. More particular examples on anti de Sitter bulks will be then considered in the next section. We will put particular emphasis on the possibility of signature-changing or null branes, but we will also compare these cases with the standard timelike branes. 

In this section, we treat the case of general 
$n$-dimensional static
spacetimes $(\N^{\pm},g^{\pm})$ (with $n> 2$)
admitting an isometry group $G_k$ of dimension $k=(n-1)(n-2)/2$ acting on the hypersurfaces 
orthogonal to the static Killing vector and containing an
isotropy group $I_s$ with $s=(n-2)(n-3)/2$ parameters.
%$SO(n-2)$ isometry group 
We will restrict to branes preserving the
$G_k$ symmetries, which leads to a symmetry-preserving matching of spacetimes, see \cite{mps}.

In appropriate adapted coordinates, the most general such spacetimes have line-elements
\begin{eqnarray*}
&&{d s^2}^+ = -A^2(r) d t^2 + B^2(r) d r^2 +
C^2(r) d \Omega^2_{\Upsilon^{n-2}_\phi},\\
&&{d s^2}^- = -\BA^2(\Br) d \Bt^{\,2} + \BB^2(\Br)
d \Br^2 + \BC^2(\Br) d \Omega_{\Upsilon^{n-2}_{\tilde{\phi}}}^2,
\end{eqnarray*}
where $d \Omega^2_{\Upsilon^{n-2}_\phi}$
is the `unit' metric on the $(n-2)$-dimensional
Riemannian space $\Upsilon^{n-2}$ of constant curvature, 
written in standard coordinates denoted by $\phi$
(and analogously for  $d \Omega_{\Upsilon^{n-2}_{\tilde{\phi}}}^2$).
The functions $A$, $B$ and $C$ depend only on $r$ and are taken to be positive
without loss of generality. The range of the coordinates $t$ and $r$ may vary from case to case, and thus it is left free in principle. The same comments apply to $\BA,\BB,\BC,\Bt$ and $\Br$.

Let us consider the $G_k$-symmetric hypersurfaces
$\S^{\pm}$ in $\N^{\pm}$. They can be defined via
$C^3$ embedding maps $\Phi_\pm : \S\rightarrow \N^{\pm}$.
%of the abstract matching hypersurface $\S$ into $(\N^+,g^+)$ 
%and $(\N^-,g^-)$, respectively. 
Taking local coordinates $\{\xi,\varphi^{M} \}$ on the abstract 
matching hypersurface $\S$
($M,N,\dots = 2 \dots n-1$), where $\{\varphi^{M}\}$ are standard
coordinates on $\Upsilon^{n-2}$, the embeddings $\Phi_{\pm}$ can
be written in local form as
\begin{eqnarray*}
&&\Phi_+(\xi,\varphi^{M} ) \equiv  \{ t=t(\xi), r=r(\xi), 
\phi^M=\varphi^M\}~~~(\S^+),\\
&&\Phi_-(\xi,\varphi^{M} ) \equiv \{ \Bt=\Bt(\xi), \Br=\Br(\xi), 
\Bphi^M=\varphi^M\}~~~(\S^-).
\end{eqnarray*}
The images under the differential maps $\d \Phi_{\pm}$ of the tangent space
basis $\{\partial_{\xi},\partial_{\varphi^M}\}$
on $\S$ are of course bases of the tangent spaces on $\S^{\pm}$.
They read explicitly
\begin{eqnarray*}
&&{\vec e_{\xi}}^{\;+}
=\left. \dot t \partial_t +
\dot r \partial_r \right|_{\S^+}, \qquad
{\vec e_{\varphi^M}}^{\;+}=\left. \partial_{\phi^M}\right|_{\S^+},\\
&&{\vec e_{\xi}}^{\;-}=\left. \dot{\Bt} \partial_{\Bt} +
\dot{\Br} \partial_{\Br} \right|_{\S^-}, \qquad
{\vec e_{\varphi^M}}^{\;-}= \left. \partial_{\tilde{\phi}^M}\right|_{\S^-},
\end{eqnarray*}
where the dot means differentiation with respect to $\xi$. 
Defining the functions 
\begin{subequations}
  \label{ffpm}
  \begin{eqnarray}
    \ff^{+} &\equiv& \left .
      -A^{2} \,\dot{t}^{2} + B ^{2} \,\dot{r}^{2} \right |_{\S^{+}},
    \\
    \ff^{-} &\equiv& \left .
      -\BA^{2} \,\dot{\Bt}^{2} + \BB^2 \dot{\Br}^{2} \right |_{\S^{-}},
  \end{eqnarray}
\end{subequations}
a simple calculation shows that the two first fundamental 
forms inherited by $\S$ from $\N^{\pm}$ coincide if and only if
\begin{equation}
\ff^+ = \ff^- \equiv \ff,  \qquad C\eqq \BC \equiv a(\xi), \label{eq:es}
\end{equation}
so that the induced metric on the brane takes the form
\begin{equation}
\fbox{$\left.{d s^2}\right|_{\S}= \ff(\xi) d\xi^2 + a^2(\xi) 
d \Omega_{\Upsilon^{n-2}_{\varphi}}^2.$}
\label{eq:ds2FLRW_N}
\end{equation}

Thus, the brane $\S$ will have in general a Lorentzian phase $\S_L$ where $N<0$, an Euclidean phase $\S_E$ defined by $N>0$, and a null phase $\S_0$ where $N=0$.
The Lorentzian part $\S_L$ describes a Robertson-Walker (RW) spacetime with $\xi$ related to the standard cosmic time $T(\xi)$ by 
\begin{equation}
\dot{T} = \sqrt{-\ff} \hspace{3mm} \mbox{on $\S_L$.}\label{cosmic}
\end{equation}
%Clearly, the
%type $k=-1,0,1$ of the FLRW cosmology depends on the curvature
%of $\Upsilon^{3}$.
The whole brane is foliated by
homogeneous and isotropic (maximally symmetric) spacelike hypersurfaces.
Changes of signature occur at given ``instants of time''
corresponding to the values $\xi_m$ of $\xi$
where $\ff$ vanishes but is not identically zero in any neighbourhood of $\xi_m$.
%such that $N|_{I_{\xi_m}}\neq 0$ for arbitrarily small intervals $I_{\xi_m}$ centered on $\xi_m$. 
The set of all such $\xi_m$ define the signature-changing set $S$ of $\S$.
%, if we assume that $\ff$ is decreasing everywhere).

From the point of view of the Lorentzian part of the brane the Lorentzian geometry becomes singular at $S$.  We shall describe later the type of singularity that any observers living on $\S_L$ will see there. We must emphasize, however, that this singularity
exists {\it only} from the {\em inner point of view} of the Lorentzian part $\S_L$, and concerns {\em only} the brane's ``Lorentzianity''.
{\em Neither the bulk nor the hypersurface $\S$ defining the
brane have any singularity anywhere} for regular functions $N(\xi)$ and $a(\xi)$.

%We assume that the signature changes
%so that $\ff$ vanishes somewhere. 
%We shall denote this value by $\xi_0$.
 
In order to complete the matching and
have a well-defined bulk and brane, we need to 
choose a rigging and solve the algebraic equations (\ref{rigg}).
For convenience we choose normal one-forms of $\S^{\pm}$ with the same norm 
at points $\Phi_{\pm}(p), p \in \S$. One possibility (not unique,
of course) is
\begin{eqnarray*}
  \bm \nany^+ &=& 
  \left.AB \left(-\dot{r}\,\d t + \dot{t}\,\d r\right)\right|_{\S^+},
  \\
  \bm \nany^-&=& 
    \left.\BA\BB \left(-\dot{\Br}\,\d \Bt +
        \dot{\Bt}\,\d \Br\right)\right|_{\S^-}.
\end{eqnarray*}
Note that $(\bm{\nany}^{+},\bm{\nany}^{+}) = (\bm{\nany}^{-},\bm{\nany}^{-})= - \ff$
provided that the preliminary matching conditions (\ref{eq:es}) hold.
A suitable rigging on $\S^{+}$ is
$$
\vec{\ell}_{+}= \epsilon_1 (- A^{-2} \dot{r} ~ \Part +
B^{-2} \dot{t} ~ \Parr ) |_{\S^{+}}
$$
where $\epsilon_1$ selects the subregion of the 
spacetime we are choosing; see subsection \ref{signs} below.
Note that $\nany^+_{\alpha} {\ell}_{+}^{\alpha} \neq 0 $ everywhere on
$\S^{+}$, as required. To find the rigging 
$\vec{\ell}_{-}$ satisfying (\ref{rigg}), observe that $\vec{\ell}_{+}$ is orthogonal
to the tangent vectors of $\Upsilon^{n-2}$, which
implies that $\vec{\ell}_{-}$ must be a linear combination of $\partial_{\Bt}$
and $\partial_{\Br}$. Thus, we can write without loss of generality
$$
\vec\ell_{-}= \epsilon_1 L ( -\alpha^{2} \BA^{-2}  \dot\Br~ 
\Partt + \BB^{-2} \dot\Bt ~ \Parrr)|_{\S^{-}}
$$
where $L\neq 0$ and $\alpha^2$ are coefficients fulfilling the
equations
% $ \epsilon_1 g^{+}_{\mu\nu} \ell_+^\mu  e_\xi^+{}^\nu=
% \epsilon_1 g^{-}_{\mu\nu} \ell_-^\mu  e_\xi^-{}^\nu$ and
% $ g^{+}_{\mu\nu} \ell_+^\mu  \ell_+^\nu =g^{-}_{\mu\nu} \ell_-^\mu  \ell_-^\nu$,
% which explicitly read, respectively,
\begin{eqnarray}
  &&\epsilon_1 g^{+}_{\mu\nu} \ell_+^\mu  e_\xi^+{}^\nu \eqq
   \epsilon_1 g^{-}_{\mu\nu} \ell_-^\mu  e_\xi^-{}^\nu:\label{eq:l1a}\\
  &&\hspace{2cm}
  2 \,\dot{r} \,\dot{t} \eqq L \left(\alpha^{2} + 1\right)\,
  \dot{\Bt} \dot{\Br}\, ,
  % =\epsilon_1 g^{-}_{\mu\nu} \ell_-^\mu  e_\xi^-{}^\nu
  \nonumber\\
  &&g^{+}_{\mu\nu} \ell_+^\mu  \ell_+^\nu \eqq
  g^{-}_{\mu\nu} \ell_-^\mu  \ell_-^\nu: \label{l.l}\\
  &&\hspace{2cm}
  - \frac{\dot{r}^2}{A^2} + \frac{\dot{t}^2}{B^2} \eqq
  L^2 \left ( - \frac{\alpha^4 \dot{\Br}^2}{\BA^2} + \frac{\dot{\Bt}^2}{\BB^2}
\right).
%=g^{-}_{\mu\nu} \ell_-^\mu  \ell_-^\nu.
\nonumber
\end{eqnarray}
The second equation involves $L$ quadratically. In order to obtain
a linear equation in $L$ which will be useful below, let us 
consider the linear combination of (\ref{l.l}) times $N$ minus the square of (\ref{eq:l1a}).
%$( \vec{\ell}_{\pm} , \vec{\ell}_{\pm} ) \ff^{\pm} -
%( \vec{\ell}_{\;\pm}, {\vec e_{\xi}}^{\;\pm})^2$,
%which obviously must take the same values on $\S^+$ and $\S^-$.
The resulting expression is a perfect square. Taking its square
root, which introduces an extra sign $\epsilon$, we get
% $\epsilon_1 \nany^+_\alpha \ell_+^\alpha  |_{\S^{+}} =
% \epsilon \epsilon_1 \nany^-_\alpha \ell_-^\alpha |_{\S^{-}}$,
% which explicitly reads
\begin{eqnarray}
  &&\epsilon_1 (\nany^+_\alpha \ell_+^\alpha)  |_{\S^{+}} \eqq
  \epsilon \epsilon_1 (\nany^-_\alpha \ell_-^\alpha) |_{\S^{-}}:\label{eq:l1b}\\
  &&\hspace{1cm}
  \frac{A}{B} \dot{t}^2+ \frac{B}{A} \dot{r}^2
  \eqq  \epsilon L \left( \frac{\BA}{\BB} \dot{\Bt}^2+\alpha^2
    \frac{\BB}{\BA} \dot{\Br}^2 \right).
  % = \epsilon \epsilon_1 \nany^-_\alpha \ell_-^\alpha |_{\S^{-}},
  \nonumber
\end{eqnarray}
Due to the positivity of the rest of the factors, this
equation readily implies that $\epsilon=\sign(L)$.
The fact that the above combinations can be
written in the covariant form (\ref{eq:l1b}) is not by chance. It simply accounts for
the a posteriori identification (after the matching is completed) of
$\bm \normal^+$ with $\bm \normal^-$ (see Section \ref{genIsrael}):
this trivially implies
$(\bm\normal^+,\bm\normal^+)=(\bm\normal^-,\bm\normal^-)$,
which thanks to choosing $\bm\nany^{+}$ and $\bm\nany^{-}$
with the same norm yields 
$( \nany^+_\alpha \ell_+^\alpha )^2 =( \nany^-_\alpha \ell_-^\alpha )^2$.
Thus, (\ref{eq:l1b}) follows for a certain sign $\epsilon$.
Moreover, 
as a result, the identification of $\bm \normal^+$ with $\bm \normal^-$
clearly leads now to the identification of
$\bm\nany^{+}$ with $\epsilon \bm\nany^{-}$.
%--- recall that both are normals
%and have the same norm --- which implies
In fact, it turns out that
the first equation in (\ref{rigg}) can be substituted by this
relation (\ref{eq:l1b}) ---whenever the
normal one-forms $\bm\nany^\pm$ have the same norm--- provided that the set $\S_0$ has empty interior.

\subsection{The energy-momentum tensor on the brane}
In order to calculate the singular part of the Einstein tensor distribution, and thereby the energy-momentum tensor on the brane, we need
to know $[\hh_{ab}]$. After a straightforward calculation using the definition
(\ref{hh}) we obtain
\begin{eqnarray}
  &&\epsilon_1\, [\hh_{\xi\xi}]=-\dot{r}\ddot{t}-\dot{t}\ddot{r}
  +L\left(\alpha^2\dot{\Br}\ddot{\Bt}+\dot{\Bt}\ddot{\Br}\right)
  \nonumber \\
  &&\qquad
  -\dot{r}^2\dot{t}\left(2\frac{A_{,r}}{A}+
    \frac{B_{,r}}{B}\right)-\dot{t}^3\frac{A A_{,r}}{B^2}\nonumber \\
  &&\qquad
\left .  +L\left[
\dot{\Br}^2\dot{\Bt}
\left ( 2\alpha^2 \frac{\BA_{,\Br}}{\BA}+ \frac{\BB_{,\Br}}{\BB} \right ) +
 \dot{\Bt}^3 \frac{\BA \BA_{,\Br}}{\BB^2} \right] \right |_{\Sigma}, \label{exprhhxixi}
  \\
  && [\hh_{MN}] \d \varphi^M \d \varphi^N =  \label{exprhh} \\
  &&\qquad
\left .  \epsilon_1\left( \dot{t}\frac{C C_{,r}}{B^2}
    -L \dot{\Bt}\frac{\BC \BC_{,\Br}}{\BB^2}\right)  \right |_{\Sigma}  \d \Omega^2_{\Upsilon^{n-2}_\phi}
  \equiv \left [ \hh \right ] \d \Omega^2_{\Upsilon^{n-2}_\phi},\nonumber \\
  && [\hh_{\xi M} ] = 0, \nonumber 
  \label{eq:hhmm}
\end{eqnarray}
where, for later convenience, we have defined
$$
[\hh] \equiv \hh^{+} - \hh^{-}
$$
with
\begin{equation}
\left.   \hh^{+} \equiv  \epsilon_1 \dot{t}\frac{C C_{,r}}{B^2} \right |_{\Sigma}
, \qquad
 \left . \hh^{-} \equiv  \epsilon_1 L \dot{\Bt}\frac{\BC \BC_{,\Br}}{\BB^2}\right |_{\Sigma}.
  \label{hhpm}
\end{equation}
Next, we must use expression (\ref{eq:tt}) to obtain the tensor $\tt_{\mu\nu}$. 
Obviously, the explicit form of this tensor depends on the coordinate system
used to describe the spacetime. 
%(this is in particular true also for the
%components of the one forms $\bm w^a$).
Since the matching procedure allows
for different coordinate systems on each side of the matching
hypersurface we need to choose one of them. For definiteness we choose the
coordinate system on $\N^{+}$. 
%We then have
%\begin{eqnarray*}
%\bm w^\xi &=& \frac{\epsilon_1}{AB (\ell_{+}^\alpha \nany^{+}_\alpha) }
%\left.\left( A^2 \dot{t}\d t+ B^2 \dot{r}\d r\right)\right|_{\S^+},\\
%\bm w^{m} &=& \left.\d \phi^m\right|_{\S^+}.
%\end{eqnarray*}
%where $\ell_{+}^{\alpha} \nany^{+}_{\alpha}$ can be read off from the first
%half of (\ref{eq:l1b}). We can now use this expression in (\ref{eq:hh})
%and substitute the result in (\ref{eq:tt})
%in order to evaluate the singular part of the Einstein
%tensor on the brane. 
Using the explicit expressions (\ref{exprhhxixi}-\ref{hhpm}) for $[\hh_{ab}]$ together with the fact that $\ell^{\mu}[\hh_{\mu\nu}]=0$, and after some calculations, the final result can be conveniently written as
\begin{eqnarray}
  \tt_{\mu\nu} dx^{\mu} dx^{\nu} &=& 
  - \frac{ \left( n-2 \right ) [ \hh ] }
  {C^2 (\ell_{+}^\alpha \nany^{+}_\alpha)^2 }
  \left ( A^2 \dot{t} dt - B^2 \dot{r} dr \right )^2
  \nonumber\\
  &&- \left .\frac{C^2 [\hh_{\xi\xi}] +
      (n-3) \ff [ \hh ] }{(\ell_{+}^\alpha \nany^{+}_\alpha)^2}
    \d  \Omega_{\Upsilon^{n-2}_{\phi}}^2 \right |_{\S^{+}} .
  \label{tensortt}
\end{eqnarray}

As expected, $\tt_{\mu\nu}$
is directly related to the quantities $[\hh_{\xi\xi}]$ and $[\hh]$. However,
expressions (\ref{exprhhxixi}) and (\ref{exprhh}) for these two quantities
are not quite satisfactory yet because
they involve $L$ and $\alpha$ which are the solutions of the algebraic
equations (\ref{eq:l1a}) and (\ref{eq:l1b}). Solving directly for $L$ and $\alpha$
and substituting into (\ref{exprhhxixi}) and (\ref{exprhh}) is not convenient
since the preliminary matching conditions must also be
taken into account.
We leave the details of this somewhat tricky calculation to the Appendix \ref{app:Hs}
and quote here the final results.
It turns out that, at points where $\ff \neq 0$, $[\hh]$ can be written in the symmetric form
\begin{equation}
  [\hh] =  
  \frac{a \left (\ell_+^{\alpha} \nany^+_{\alpha} \right )}{\ff} 
  \left . \left (\epsilon \frac{\BA}{\BB} \BC_{,\Br} \dot{\Bt} - 
      \frac{A}{B} C_{,r} \dot{t} \right) \right |_{\S}
  \label{finalhh}
\end{equation}
while $[ \hh_{\xi\xi} ]$ reads
\begin{eqnarray}
  &&\frac{1}{\left (\ell_{+}^{\alpha} \nany^{+}_{\alpha} \right )}
  \left [ H_{\xi\xi} \right ] \dot{t} \dot{\Bt} =
  \nonumber \\
  &&\qquad %\left . 
  \epsilon \dot{t} \left[ \frac{\BA_{,\Br}}{\BB} \dot{\Bt}^2 + 
    \frac{\BB_{,\Br}}{\BA} 
    \dot{\Br}^2 - \frac{\BB}{\BA} \left (\frac{\dot{\ff}}{2\ff} \dot{\Br} - 
      \ddot{\Br} \right ) \right] 
  \nonumber \\
  &&\qquad
  \left .- \dot{\Bt} \left[ \frac{A_{,r}}{B} \dot{t}^2 + \frac{B_{,r}}{A} 
      \dot{r}^2 - \frac{B}{A} \left (\frac{\dot{\ff}}{2\ff} \dot{r} - \ddot{r}
      \right ) \right] \right |_{\S}.
  \label{finalhhxixi}
\end{eqnarray}
Due to the presence of $\ff$ in the denominator it may seem at first sight
that the expressions (\ref{finalhh}, \ref{finalhhxixi}) diverge when we
approach the null phase $\S_0$. This is however not the case
because $[\hh_{\mu\nu}]$ is by construction well-defined everywhere on $\Sigma$. 
This also follows directly from expressions (\ref{exprhhxixi}) and (\ref{exprhh}), which are
regular on $\S_0$.
% \mnote{\marc No crec necessary demostrar que el limit
% existeix, quan ja se sap. Si es vol fer, caldria evitar
% els signes $\sigma$ i $\tilde{\sigma}$, que son differents
% als que s'introdueixen despres}

For completeness, let us include here an expression
for $[\hh]$ at points on $\S_0$. Equations (\ref{ffpm}) become 
$$\dot{r}^2|_{\S_0} =\left. \frac{A^2}{B^2}
\dot{t}^2\right|_{\S_0}, \hspace{3mm}
\dot{\Br}^2 |_{\S_0}= \left.\frac{\BA^2}{\BB^2} \dot{\Bt}^2\right|_{\S_0}.
$$
This implies that neither $\dot{r}, \dot{t}$, $\dot{\Br}$ nor $\dot{\Bt}$
can vanish on $\S_0$ (otherwise $\Phi_{\pm}$ would not be embeddings).
Then,
equation (\ref{l.l}) implies that 
$$\alpha^2|_{\S_0} =1,$$ which inserted in 
(\ref{eq:l1a}) gives
\begin{equation}
\label{eq:LaS0}
L|_{\S_0} = \left.\frac{\dot{r} \dot{t}}{\dot{\Br} \dot{\Bt} }\right|_{\S_0}.
\end{equation}
Using all this in (\ref{exprhh}) and recalling $\dot{a}= C_{,r} \dot{r} |_{\S}
= C_{,\Br} \dot{\Br} |_{\S}$, 
we finally obtain
\[
[\hh]|_{\S_0}= \sign(\dot{r} \dot{t})  a \dot a \frac{\epsilon_1}{AB}
\left.\left( 1 - 
\frac{\dot r^2 B^2}{\dot \Br^2 \BB^2}\right)\right|_{\S_0}.
\]
%which is regular on $S$.
%This shows that (\ref{Singular?}), as well as (\ref{finalhhxixi}), are regular at the null phase $\S_0$ of $\S$, even at the set $S$ where $\S$ changes charatacter.

Once we have computed the singular part $\tt_{\mu\nu}$ of the Einstein tensor, given by (\ref{tensortt}), the energy-momentum tensor on the brane follows directly from 
(\ref{eq:brane_equation}). A convenient way of describing this object is
via its eigenvalues.
Since $\tt_{\mu\nu} n^{\mu} = 0$ holds identically, the
rank of the tensor $\tt_{\mu\nu}$,
and hence of $\ttt_{\mu\nu}$ is at most $n-1$ and $0$
is always one of its eigenvalues.
In order to evaluate the remaining eigenvalues of $\ttt_{\mu\nu}$ ---
which correspond to the eigenvalues of $\ttt_{ab}$ wherever
$\S$ is not null--- let us rewrite (\ref{tensortt}) as
\begin{equation}
\ttt_{\mu\nu}dx^{\nu} dx^{\mu} \eqq - \ff^{-1}
\hat\varrho
 ( A^2 \dot{t} dt - B^2 \dot{r} dr )^2 + \hat p a^2
d\Omega_{\Upsilon^{n-2}_{\phi}}^2 \label{ttthatphatrho}
\end{equation}
where we have defined
\begin{eqnarray*}
  \kappa_n^2 \hat\varrho 
  &\equiv& 
%- \frac{1}{\ff} (\tt_{\mu\nu}+\tension g_{\mu\nu}){e_{\xi}}^{\mu} {e_{\xi}}^{\nu}=
  \left.\frac{ (n-2) \ff[\hh]}{a^2
    \left(\ell_{+}^{\alpha}\nany^{+}_{\alpha}\right)^2}\right |_{\S}
  \\
  &=& \left.
    \frac{(n-2)}{a\left(\ell_{+}^{\alpha} \nany^{+}_{\alpha}\right)} 
    \left (\epsilon \frac{\BA}{\BB} \BC_{,\Br} \dot{\Bt} - 
      \frac{A}{B} C_{,r} \dot{t} \right) \right |_{\S},\\
%\end{eqnarray*}
%\begin{eqnarray*}
  \kappa_n^2 \hat p  &\equiv& 
  - \frac{ a^2 [ \hh_{\xi\xi} ] + (n-3)
    \ff [\hh]}{a^2 \left( \ell_{+}^{\alpha} \nany^{+}_{\alpha} \right)^2}
  \\
  &=&- \frac{[ \hh_{\xi\xi} ]}
  { \left(\ell_{+}^{\alpha}\nany^{+}_{\alpha} \right)^2}
  - \frac{n-3}{n-2} \kappa_n^2 \hat\varrho .
\end{eqnarray*}
Since the one-form $-A^2 \dot{t} dt + B^2 \dot{r} dr$
appearing in (\ref{ttthatphatrho}) is precisely the tangent vector $\vec{e}_{\xi}$ with index down
and that its norm is simply $\ff$, it follows easily 
that the remaining eigenvalues of
$\ttt_{\mu\nu}$ are precisely $-\hat{\rho}$ and $\hat{p}$.

The explicit expression for $\hat p$ can be read off directly
from the previous formula and the use of (\ref{finalhhxixi}).
However, it is simpler and more convenient to note the following identity
which follows after a straightforward, if somewhat long, calculation
% \begin{eqnarray}
%   \dot{\hat\varrho} +
%   \left(\frac{(\ell_{+}^{\alpha}\nany^{+}_{\alpha})\dot{}}
%     {(\ell_{+}^{\alpha}\nany^{+}_{\alpha})}-
%     \frac{\dot{\ff}}{2 \ff} \right)\hat\varrho +
%   \left (n-2 \right )\frac{\dot{a}}{a} 
%   \left (\hat\varrho + \hat p \right)
%   \nonumber\\
%   + \frac{ 
%     (n-2)}{\kappa_n^2 (\ell_{+}^{\alpha}\nany^{+}_{\alpha})}
%   \nonumber\\
%   \times\left[ \frac{A^2 \dot{t} \dot{r} }{a}
%     \left ( \frac{C_{,r}}{AB} \right )_{,r}  -
%     \epsilon \frac{\BA^2 \dot{\Bt} \dot{\Br} }{a}
%     \left ( \frac{\BC_{,\Br}}{\BA \BB} \right )_{,\Br} \right]
%   \nonumber\\
%   = 0. 
%   \label{conser0}
% \end{eqnarray}
\begin{widetext}
\begin{equation}
  \dot{\hat\varrho} +
  \left(\frac{(\ell_{+}^{\alpha}\nany^{+}_{\alpha})\dot{}}
    {(\ell_{+}^{\alpha}\nany^{+}_{\alpha})}-
    \frac{\dot{\ff}}{2 \ff} \right)\hat\varrho +
  \left (n-2 \right )\frac{\dot{a}}{a} 
  \left (\hat\varrho + \hat p \right)
  %\nonumber\\
  + \frac{ 
    (n-2)}{\kappa_n^2 (\ell_{+}^{\alpha}\nany^{+}_{\alpha})}
  \left \{ \frac{A^2 \dot{t} \dot{r} }{a}
    \left ( \frac{C_{,r}}{AB} \right )_{,r}  -
    \epsilon 
    \frac{\BA^2 \dot{\Bt} \dot{\Br} }{a}
    \left ( \frac{\BC_{,\Br}}{\BA \BB} \right )_{,\Br} \right \}
  %\nonumber\\
  = 0. 
  \label{conser0}
\end{equation}
\end{widetext}
This identity clearly resembles a continuity equation. We shall
see that this is exactly the case, with explicit applications for anti-de Sitter bulks.

\subsection{The meaning of the signs}
\label{signs}
Since our convention is
that the rigging $\vec\ell_{+}$
of $\S^{+}$ points towards the submanifold $\M^{+} \subset \N^{+}$ and that
$\vec \ell_{-}$ points outwards from the submanifold $\M^{-} \subset
\N^{-}$ it follows that 
choosing the signs 
$\epsilon$ and $\epsilon_1$ amounts to selecting which subsets $\M^{\pm} \subset
\N^{\pm}$ are taken to perform the matching. 
%From the results in \ref{subsub},
%not all possible orientations will be possible in the case of a 
%boundary hypersurface for $\M^{\pm}$ with non-empty null phase, that is, 
%when the function $\ff$ vanishes somewhere on the boundary hypersurface. 

Hitherto everything is valid for general branes. However, 
if $\S$ is non-null everywhere, 
%(and thus its signature is constant),
the algebraic equations (\ref{eq:l1a},\ref{l.l})
admit two different solutions for $L$ for each choice of
$\vec{\ell}_{+}$, and these two solutions have a different sign $\epsilon$,
according to (\ref{eq:l1b}). On the other hand, if there is a point $p$ where $\S$ becomes null,
from Lemma \ref{uniquenessNull}
there is at most one solution for $\epsilon$.
%This fact can be checked explicitly in equations (\ref{eq:l1a},\ref{l.l})
%by noticing that the resulting \emph{a priori} quadratic
%equation for $L$ (with no $\alpha$) contains the factor $\ff$ in the $L^2$ term,
%and thus is linear in $L$ at $p$. 
Let us determine its value. We already know that $\epsilon = \sign (L)$, but
$L|_{\S_0}$ has been already computed on (\ref{eq:LaS0}),
% On $\S_0$, 
% $$\dot{r}^2|_{\S_0} =\left. \frac{A^2}{B^2}
% \dot{t}^2\right|_{\S_0}, \hspace{3mm}
% \dot{\Br}^2 |_{\S_0}= \left.\frac{\BA^2}{\BB^2} \dot{\Bt}^2\right|_{\S_0},
% $$
% which implies that neither $\dot{r}, \dot{t}$, $\dot{\Br}$ nor $\dot{\Bt}$
% can vanish there (otherwise $\Phi_{\pm}$ would not be embeddings).
% Then, 
% %relations (\ref{sigmas}) hold,
% %we have $\dot{r}^2\eqqS \frac{A^2}{B^2}
% %\dot{t}^2$ and $\dot{\Br}^2 \eqqS \frac{\BA^2}{\BB^2} \dot{\Bt}^2$.
% %This introduces two signs, $\sigma$ and $\tilde\sigma$, defined by
% %\[
% %\dot{r} \eqqS \sigma \frac{A}{B} \dot{t},~~~
% %\dot{\Br} \eqqS \tilde\sigma \frac{\BA}{\BB} \dot{\Bt}.
% %\]
% %and furthermore 
% equation (\ref{l.l}) implies that 
% $$\alpha^2 =1,$$ which inserted in 
% (\ref{eq:l1a}) gives
% $$
% L = \frac{\dot{r} \dot{t}}{\dot{\Br} \dot{\Bt} } $$
and consequently 
\begin{equation}
\epsilon = \sign (L) = \sign (\dot{r} \dot{t} \dot{\Br} \dot{\Bt}) \hspace{1cm} (\Sigma_{0} \neq \emptyset)\, .\label{sign}
\end{equation}
% %\sigma\tilde\sigma A\BB\dot t^2/( B\BA \dot\Bt^2)\right|_{\S_0}
% %$$
% %(note that $\dot{t}$, $\dot{\Bt}$, $\dot{r}$ and $\dot{\Br}$ are
% %all non-zero at $\S_0$, as follows from (\ref{sigmas}).) 
Therefore, if $\S_0$ is not empty then $\epsilon$ is unique and
explicitly determined by the two embeddings. 
%the sign of $L$ is 
%$$
%\epsilon=\sigma\tilde\sigma , \,\,\,\, \mbox{if} \,\, \S_0\neq \emptyset
%$$
%which is a {\em fixed} sign given 
Since in the purely Lorentzian (or Euclidean) case $\epsilon$ is
free, we shall also keep $\epsilon$ free in order to 
compare our general results with previous works on Lorentzian branes.

With regard to the remaining sign $\epsilon_1$, 
this has not been fixed so far. 
Observe that
% this sign is that of $\ell_{+}^{\alpha}\nany^{+}_{\alpha}$, i.e.
$\epsilon_1=\sign(\ell_{+}^{\alpha}\nany^{+}_{\alpha})$,
as follows from (\ref{eq:l1b}) and the fact that $A,B,C$ have been chosen
to be positive.
%In Proposition \ref{OnlyOneWorks} we proved
%that, given a manifold-with-boundary whose boundary
%has a non-empty $\S_0$ and two different regions of a second spacetime
%separated by another boundary with non-empty null phase, there is {\it at most} one
%region of the second spacetime that can be matched to the first
%spacetime-with-boundary.  
The interpretation of this sign is, therefore, as follows. 
In the construction above, we use {\it two} spacetimes, each of which contains
a hypersurface that separates each spacetime into
two regions. So we have {\it four} regions to play with. Fixing one of
the regions in one spacetime, this may be matchable to none, one or both of the regions in the second
spacetime ---if $\S$ has a non-empty null phase, there is at most one possibility as follows
from Proposition \ref{OnlyOneWorks}. 
But, can the left-out region of
the first spacetime be matched to any of the regions in the
second? The answer is yes {\em if the originally chosen region in the first spacetime was matchable} to one of the regions in the second; and actually the region that now matches with it is
precisely the {\it complementary} part of the one that matched with the
first region of the first spacetime. In short, given two matchable spacetimes
there  {\it always} are two complementary matchings, as discussed
in detail in \cite{FST}. 
This provides an interpretation for
$\epsilon_1$: it selects which region at both sides of $\S$ in the first
spacetime is taken to perform the matching.
A scheme of the four possible different cases discussed in this
paragraph for the particular case of AdS bulks
is shown in Figure \ref{fig:casos}.

%It is worth noticing that despite the appearance of $\ff$  in the denominator
%expression (\ref{finalhhxixi}) is regular everywhere, even in the
%case of signature changing branes. This is best 
%seen by noticing that the object $X$ introduced above
%as well as $\epsilon_1 (\ell^\alpha n_\alpha)$ are, by construction,
%everywhere regular on $\S$, and thus
%the same must hold for
%$\epsilon_1 X/(\ell^\alpha n_\alpha)$,
%whose equivalent expression is $\ff^{-1} \left(
%B A^{-1} \dot{\Bt} \dot{r} - \epsilon \BB \BA^{-1} \dot{t} \dot{\Br}
%\right)$, even at points where $\ff$ vanishes. In particular the term
%within parentheses must vanish whenever $\ff$ vanishes. 

%DIDN'T HAVE TIME TO WRITE THE PROOF, I WILL COMPLETE THIS.....

\section{Signature changing branes in A\lowercase{d}S${}_n$ bulks}
\label{AdSect}
Let us now specialise to the case where $\N^{+}$  and $\N^-$ are anti-de Sitter
spaces of dimension $n$, usually denoted by AdS$_{n}$. For that, we
%rename the radial coordinates as
%$r \rightarrow \rho$ and $\Br \rightarrow \Brho$ and
choose the metric functions to be
\begin{eqnarray}
  &&A^2 = B^{-2} = k + \lambda^2 r^2, \qquad C = r, \label{1}\\
  &&\BA^2 = \BB^{-2} = k + \Blambda^2 \Br^2,  \qquad \BC = \Br,\label{2}
\end{eqnarray}
where $\lambda$ and $\Blambda$ are non-negative constants
related to the cosmological constant by means of $2\Lambda_n =-(n-1)(n-2)\lambda^2$, and analogously for the tilded ones. Here $k=-1,0,1$ corresponding to three possible coordinate systems to describe the AdS$_{n}$ spacetime. $k$ 
coincides with the sectional curvature of
%curvature for the constant curvature part of the line-elements, i.e. for 
$d\Omega_{\Upsilon^{n-2}_\phi}$.
%Eventually, the bulk will be taken to be anti-de Sitter spacetime,
%AdS${}_{5}$, for which $\ttm_{\mu\nu}=0$. There are three
%suitable coordinate systems for 
%in which the metric can be put to the form \cite{?}
%\begin{equation}
%ds^2=-(k+\lambda^2 r^2)dt^2+(k+\lambda^2 r^2)^{-1}d r^2+ r^2\doo,
%\label{AdS}
%\end{equation}
%where $\doo=d \chi^2+ \f^2_k(\chi)(d\theta^2+\sin^2\theta)d\phi^2$,
%for $\f_{1}(\chi)=\sin\chi$, $\f_{0}(\chi)=\chi$, $\f_{-1}(\chi)=\sinh\chi$,
%is the metric of $S^3$
%and $\lambda$ is a positive constant
%related to the negative cosmological constant of the spacetime by
%$\Lambda_5 =-6 \lambda^2$. 

The case of a flat bulk is included here for the values
$\lambda = 0$ and $k=1$.
%Note that
%the coordinates for which $k=0,1$ cover all the spacetime
%except at $r=0$, but
%that is not the case when $k=-1$.
When $k=0,1$, the ranges of the
non-angular coordinates are $-\infty <t<\infty$ and $r> 0$,
the center of symmetry being located at 
%and they never cover
%the whole AdS${}_{n}$ spacetime except for 
%the center of symmetry 
$r=0$. 
In the $k=-1$ case, though, the range of $r$ is further restricted to
$r > 1/\lambda$.
%and does not cover the whole of AdS${}_{n}$.

Due to Corollary \ref{res:2} we cannot construct the bulk by
gluing together 
two copies of a submanifold-with-boundary of AdS${}_{n}$ if the boundary has a non-empty null phase. However, there is no a priori
obstruction to consider two different submanifolds-with-boundary of AdS${}_{n}$
or, more generally, to try and paste a region of AdS${}_{n}$ with
another region of a possibly different anti-de Sitter space, $\widetilde{\mbox{AdS}}_n$,
with another cosmological constant.
% $\widetilde{\Lambda_n} = -(n-1)(n-2) \tilde{\lambda}^2/2$.
%The line-element of $\widetilde{\mbox{AdS}}_5$ will be written as
%\[
%d\tilde{s}^2=-(k+\tilde{\lambda}^2 \trho^2)d\tilde{t}^2+
%(k+\tilde{\lambda}^2 \trho^2)^{-1}d\trho^2+\trho^2\doo.
%\]
For simplicity, and as in the previous section, we will only consider 
branes $\S$ with spherical, plane or hyperboloidal
symmetry. 
%In ohter  words, the
%boundary is defined locally by functions $F(t,r)=0$ and
%$\tilde{F}(\tilde{t},\trho)=0$). 

%(Let us recall that for $k=-1$ we must restrict ourselves to
%$r>1/\lambda$.)
Particularizing the equations of the previous section to the explicit functions (\ref{1}-\ref{2}),
we get
\begin{equation}
a(\xi) = r(\xi) = \Br(\xi) \label{artilder}
\end{equation}
while (\ref{ffpm}) and (\ref{eq:es}) yield ordinary
differential equations for $t(\xi)$ and $\Bt(\xi)$ in terms of $\ff(\xi)$
\begin{subequations}
  \label{eq:ts}
  \begin{eqnarray}
    \dot{t} &=& \frac{\sigma a}{k + \lambda^2 a^2} \sqrt{ \frac{\dot{a}^2}{a^2}
      - \ff \left ( \frac{k}{a^2} + \lambda^2 \right) }, \\
    \dot{\Bt} &=& \frac{\tilde\sigma a}{k + \Blambda^2 a^2}
    \sqrt{ \frac{\dot{a}^2}{a^2}
      - \ff \left ( \frac{k}{a^2} + \Blambda^2 \right) }
  \end{eqnarray}
\end{subequations}
where $\sigma$ and $\tilde\sigma$ are two signs. 
%in direct correspondence with those introduced in (\ref{sigmas}).
% (note that it has been chosen so that it
% coincides with the $\sigma$ introduced above whenever $\dot{a} >0$,
% although that is not required).
For compactness, it is convenient
to define 
%the following relevant combination of signs
\begin{equation}
\varepsilon\equiv \epsilon \sigma \tilde\sigma, \label{varep}
\end{equation}
which will in fact substitute $\tilde\sigma$.
%From the discussion in subsection \ref{signs} and using (\ref{artilder}) 
%we know that, if $\S$ is to admit a
%non-empty null phase $\S_0$, then $\varepsilon=1$ necessarily.

With these expressions
we can write down the explicit form for $\hat\varrho$ and $\hat p$
in the present case:
\begin{widetext}
\begin{eqnarray}
\frac{\kappa^2_n \;\hat\varrho}{n-2}&=&\frac{\sigma}
{(\ell_{+}^{\alpha}\nany^{+}_{\alpha})}
\left (\varepsilon
\sqrt{\frac{\dot{a}^2}{a^2}-\ff \left (\frac{k}{a^2}+\Blambda^2 \right)}
- \sqrt{\frac{\dot{a}^2}{a^2}- \ff \left( \frac{k}{a^2}+\lambda^2
\right)} \right ) , \nonumber\\
\kappa_n^2(\ell_{+}^{\alpha}\nany^{+}_{\alpha})
\left (\hat p + \frac{n-3}{n-2} \hat\varrho \right) &=& 
\varepsilon\sigma
\frac{\left ( \Blambda^2 \ff + \frac{\dot{\ff}}{2\ff} \frac{\dot{a}}{a}
- \frac{\ddot{a}}{a} \right )}{
\sqrt{\frac{\dot{a}^2}{a^2}-\ff \left (\frac{k}{a^2}+\Blambda^2 \right)}}
- \sigma\frac
{\left ( \lambda^2 \ff + \frac{\dot{\ff}}{2\ff} \frac{\dot{a}}{a}
- \frac{\ddot{a}}{a} \right )}{
\sqrt{\frac{\dot{a}^2}{a^2}-\ff \left (\frac{k}{a^2}+\lambda^2 \right)}},
\nonumber
\end{eqnarray}
\end{widetext}
where
\[
\ell_{+}^{\alpha}\nany^{+}_{\alpha}=
\epsilon_1\left(2 \frac{\dot a^2}{k+a^2\lambda^2}-\ff\right).
% \frac{\epsilon_1 a^2}{1+\lambda^2 a^2}
% \left[2\frac{\dot a^2}{a^2}-\ff\left(\frac{1}{a^2}+\lambda^2\right) \right].
\]
Regarding the identity (\ref{conser0}), it simplifies to
\begin{equation}
\dot{\hat\varrho}+
\frac{d}{d\xi}
\left(
\log\frac{|\ell_{+}^{\alpha}\nany^{+}_{\alpha}|}
{\sqrt{|N|}}
\right) %^{\bullet}
%\left(\frac{(\ell_{+}^{\alpha}\nany^{+}_{\alpha})\dot{}}
%{(\ell_{+}^{\alpha}\nany^{+}_{\alpha})}-
%\frac{\dot{\ff}}{2 \ff} \right)
\hat\varrho+
(n-2)\frac{\dot{a}}{a}(\hat\varrho+\hat p)=0.
\label{eq:pre-cons}
\end{equation}
At points where $\ff \neq 0$ (i.e. outside the null phase $\S_0$) we
can define
\begin{equation}
{\varrho} \equiv
\hat \varrho~ \frac{|\ell_{+}^{\alpha}\nany^{+}_{\alpha}|}{\sqrt{|\ff|}},
\qquad
%and
{p}\equiv
\hat p ~ \frac{|\ell_{+}^{\alpha}\nany^{+}_{\alpha}|}{\sqrt{|\ff|}},\label{1/N}
\end{equation}
so that the conservation law is obtained from
(\ref{eq:pre-cons}) in its standard form
\begin{equation}
\dot{{\varrho}}+(n-2)\frac{\dot{a}}{a}({\varrho}+{p})=0.
\end{equation}

%REASONS FOR THE NORMALISATION:

This choice of normalisation may seem artificial, but it corresponds
precisely to the choice of the unit normal vector as the rigging vector on $\S_L$.
%Thereby one can recover the usual 
%extrinsic curvature $K_{ab}$,
%the Israel formula and hence, the usual 
%energy-momentum tensor on the  brane.
%Indeed, when $\ff \neq 0$ we choose
%\[
%\vec \ell_+ =
%- \frac{\epsilon_1~ \sign(\ff)}{\sqrt{|\ff|}} \vec \nany^+,
%\qquad
%\vec \ell_- =
%- \frac{\epsilon~\epsilon_1~ \sign(\ff)}{\sqrt{|\ff|}} \vec \nany^-,
%\]
%so that all the relations above between $\vec\ell_\pm$ and $\bm\nany^\pm$
%hold together with
%\begin{eqnarray*}
%&&(\ell_{+}^{\alpha}\nany^{+}_{\alpha})=\epsilon
%(\ell_{-}^{\alpha}\nany^{-}_{\alpha})=
%\epsilon_1 \sqrt{|\ff|}\\
%&&\qquad\qquad
%\Longrightarrow~~
%\frac{|\ell_{+}^{\alpha}\nany^{+}_{\alpha}|}{\sqrt{|\ff|}}=
%\frac{|\ell_{-}^{\alpha}\nany^{-}_{\alpha}|}{\sqrt{|\ff|}}=1,
%\end{eqnarray*}
%noting that $\sign(\ff)=-1$ for the Lorentzian part of $\S$.
Therefore, ${\varrho}$ and ${p}$ are functions
%defined all over a changing character $\S$
that correspond to \emph{the energy density and pressure measured within
the Lorentzian part $\S_L$ of the brane $\S$.}
These non-hatted functions are then relevant physical quantities
one has to analyse.

To start with, recall that $\hat \varrho$ and $\hat p$
\emph{are regular} everywhere on $\S$. Thus, from (\ref{1/N}) one could be misled to think that the energy density ${\varrho}$ and pressure ${p}$ blow up when approaching a change of signature $S\cap \overline{\Sigma_{L}}$. Nevertheless,
we are going to prove in what follows that, actually, ${\varrho}$
{\em vanishes} at the signature change, and that $p$ can also be regular in many
cases; see subsection \ref{tope}.
% ((Of course, this will happen because
% $\hat \varrho$ and $\hat p$ will vanish whenever $N=0$.))

To show this and to compare with previous works on purely Lorentzian
branes in AdS (see e.g. \cite{derdol}),
let us perform the change (\ref{cosmic}) from the timelike coordinate $\xi$  to the cosmic time $T$,
%in (\ref{eq:ds2FLRW_N}) ,
%\[
%\dot T(\xi)=\sqrt{-N(\xi)},
%\]
which is suitable at points where $N< 0$,
%, i.e. at the Lorentzian part of $\S$, 
so that the line-element (\ref{eq:ds2FLRW_N}) reads on $\S_L$
\[
\left.{d s^2}\right|_{\S_L}= - d T^2 +
a^2 d \Omega_{\Upsilon^{n-2}_{\varphi}}^2.
\]
Using the notation $' = d/dT$ we have
\begin{equation}
  \label{eq:aT}
  a'=\frac{\dot a}{\sqrt{-N}},\qquad
  a''=-\frac{\ddot a}{N}+\dot a \frac{\dot N}{2 N^2},
\end{equation}
which we use to obtain
\begin{eqnarray}
&&\varrho'+(n-2)\frac{{a'}}{a}({\varrho}+{p})=0,
\label{eq:cont_L}\\
&&\frac{\kappa^2_n}{n-2}\varrho=
\nonumber\\
&&
=\sigma\epsilon_1
\left [\varepsilon
\sqrt{\frac{{a'}^2}{a^2}+ \frac{k}{a^2}+\Blambda^2 }
- \sqrt{\frac{{a'}^2}{a^2}+  \frac{k}{a^2}+\lambda^2} \right ].
\label{eq:Friedmann_L}
\end{eqnarray}
% Defining
%  \[
%  \rho\equiv \frac{\kappa_n^2}{n-2}\varrho, \qquad \text{ and } \qquad
%  H\equiv \frac{a'}{a},
%  \]
Defining the Hubble function $H\equiv a'/a$ as usual,
equation (\ref{eq:cont_L}) yields 
%the following expression for $p$:
\begin{eqnarray}
  \label{eq:p_maca}
  p&=&\varrho \left[\frac{\varepsilon}{n-2}
    \left(H'-\frac{k}{a^2}\right)
  \left(H^2 %\frac{{a'}^2}{a^2}
    + \frac{k}{a^2}+\Blambda^2\right)^{-1/2}\right.
\nonumber\\
&&\left.\times
  \left(H^2 %\frac{{a'}^2}{a^2}
    + \frac{k}{a^2}+\lambda^2\right)^{-1/2}-1
\right].
\end{eqnarray}

Passing any of the square roots of (\ref{eq:Friedmann_L}) to the left
and squaring we obtain the following respective two expressions
\begin{eqnarray}
&&\varrho~\sqrt{\frac{{a'}^2}{a^2}+ \frac{k}{a^2}+\Blambda^2 }=\nonumber\\
&&~~~\sigma\epsilon_1\varepsilon\frac{(n-2)}{2\kappa_n^2}\left(\Blambda^2-\lambda^2+\frac{\kappa_n^4}{(n-2)^2}\varrho^2\right),\label{sigma1}\\
&&\varrho~\sqrt{\frac{{a'}^2}{a^2}+ \frac{k}{a^2}+\lambda^2 }=\nonumber\\
&&~~~ -\sigma\epsilon_1\frac{(n-2)}{2\kappa_n^2}\left(\lambda^2-\Blambda^2+\frac{\kappa_n^4}{(n-2)^2}\varrho^2\right).\label{sigma2}
\end{eqnarray}
Now, squaring any of these two expressions, and provided $\varrho\neq 0$, we obtain
%Moreover, equation (\ref{eq:Friedmann_L}) implies
the following condition
\begin{eqnarray}
  &&\frac{a'^2}{a^2}+\frac{k}{a^2}=\nonumber \\
  &&~~~\frac{(n-2)^2}{4 \kappa_n^4\varrho^2}\left[
    \left(\Blambda^2+\lambda^2-\frac{\kappa_n^4}{(n-2)^2}\varrho^2 \right)^2
    -4 \Blambda^2 \lambda^2
  \right],
  \label{eq:Friedmann_brane}
\end{eqnarray}
which is usually referred to as ``the modified Friedmann equation''
for %``asymmetric''
brane-world cosmologies. 

Let us discuss these relations (\ref{eq:cont_L}-\ref{eq:Friedmann_brane}) in detail.
\begin{itemize}
\item An important remark is that (\ref{eq:cont_L}-\ref{eq:Friedmann_brane}) hold only on $\S_L$.
\item Equation (\ref{eq:cont_L})
is the usual continuity equation in $(n-1)$-dimensional RW spacetimes. The traditional 4-dimensional case is recovered by assuming $n=5$, that is, a 5-dimensional bulk.
\item Equation (\ref{eq:p_maca}) can be regarded as a Raychaudhuri-like
equation on the brane.
\item Concerning (\ref{eq:Friedmann_L}), let us first of all stress the fact that 
the modified Friedmann equation (\ref{eq:Friedmann_brane}),
which is the equation usually found in the literature as a
consequence of using the SMS \cite{SMS} formalism,
%\mnote{\raul He canviat això. No deiem res del SMS, del que no poden
%treure tota la informació}
%included for instance in references
%\cite{gogb1,derdol,gergely1},
is just one of its consequences. In other words, (\ref{eq:Friedmann_brane}) is only a \emph{necessary} quadratic condition, and its solutions still have to
satisfy (\ref{eq:Friedmann_L}). Thus, {\em the truly relevant equation, containing all the information, 
is (\ref{eq:Friedmann_L})}.
%NOTE THAT THE EQUATION FOUND BY USING SMS CORRESPONDS TO THE QUADRATIC
%EXPRESSIONS HERE. BUT THE GOOD EQUATIONS ARE THE ONES WITH SQRTs.

To see this in more detail, and its consequences,
let us focus on (\ref{sigma1}) and (\ref{sigma2}).
% notice that (\ref{eq:Friedmann_brane}) easily implies, 
% provided $\varrho \neq 0$,
%after some trivial algebra
% \begin{eqnarray}
% \sqrt{\frac{{a'}^2}{a^2}+ \frac{k}{a^2}+\Blambda^2 }=\hspace{1cm}\nonumber\\
% \frac{\sigma_1(n-2)}{2\kappa_n^2\varrho}\left(\Blambda^2-\lambda^2+\frac{\kappa_n^4}{(n-2)^2}\varrho^2\right),\label{sigma1}\\
% \sqrt{\frac{{a'}^2}{a^2}+ \frac{k}{a^2}+\lambda^2 }=\hspace{1cm}\nonumber\\
% \frac{\sigma_2(n-2)}{2\kappa_n^2\varrho}\left(\lambda^2-\Blambda^2+\frac{\kappa_n^4}{(n-2)^2}\varrho^2\right)\label{sigma2}
% \end{eqnarray}
% where $\sigma_1,\sigma_2$ are the signs such that the left-hand side square roots are positive. By introducing them
% into (\ref{eq:Friedmann_L}) and after some algebra the following values are found
%\mnote{\marc Calculation with reduce. No need of imposing $\rho>0$ nor $\lambda^2 - \tilde{\lambda}^2 \neq 0$} 
%the following restrictions are found provided $\varrho >0$: for general values of $\lambda,\Blambda$
%$$
%\varepsilon\sigma_1-\sigma_2=2\sigma\epsilon_1
%$$
%and if in addition $\lambda^2-\Blambda^2\neq 0$, then also
%$$
%\sigma_1=\varepsilon\sigma\epsilon_1, \hspace{3mm} \sigma_2=-\sigma\epsilon_1. 
%$$
%Note, however, that at least one of the signs $\sigma_1,\sigma_2$ is equal to 1 ---which particular one depends on the sign of $\lambda^2-\Blambda^2$. This restricts in turn the feasible values of $\sigma\epsilon_1$ and $\varepsilon$, and also implies that\mnote{\marc I don't get the strict inequalities. Also I only get $|\rho|$. If old version true,
%we should explain how they are found} 
By multiplying these two equations we obtain
\[
0\leq -\varepsilon\left(\left(\frac{\kappa_n^2}{n-2}\varrho\right)^4
-(\Blambda^2-\lambda^2)^2\right),
\]
from where
\begin{eqnarray}
&&\frac{\kappa_n^2}{n-2} |\varrho| \leq  \sqrt{|\Blambda^2-\lambda^2|}\,\,\,\, \mbox{if}\,\,\,\, \varepsilon =1,  \label{limit1}\\
&&\frac{\kappa_n^2}{n-2} |\varrho| \geq \sqrt{|\Blambda^2-\lambda^2|}\,\,\,\, \mbox{if}\,\,\,\, \varepsilon =-1 .\label{limit2}
\end{eqnarray}
%Also, demanding $\varrho>0$ we have from (\ref{sigma1}) and (\ref{sigma2})
%that
%$\Blambda^2-\lambda^2<0\Rightarrow -\sigma\epsilon_1=1$, whereas
%$\Blambda^2-\lambda^2>0\Rightarrow \sigma\epsilon_1 \varepsilon=1$,
%and that $\Blambda=\lambda \Rightarrow \varepsilon=\sigma\epsilon_1=-1$.

\end{itemize}

%\mnote{
%\marc No entenc el paragraf. En particular el ``however'' em descoloca.
%\raul Ho he separat de la llista, aviam ara...}
The important expression (\ref{eq:Friedmann_L}) appears in
full form in \cite{AABS,AS}, and partially in
%---and generalizes relations found in
\cite{gogb1,derdol} %---
for the so-called `shell cosmologies'.

In expression (\ref{eq:Friedmann_L}) (for $n=5$), {\em both}
the usual `brane', i.e. using $Z_2$-mirror symmetry in AdS$_5$, and the `shell'
cosmologies are naturally recovered. The $Z_2$-mirror branes require $\lambda=\Blambda$ and, as we already know, are incompatible with signature changes or null phases. The shell cosmologies,
also referred to as `asymmetric' brane cosmologies in \cite{gergely1}, require on the contrary that $\lambda\neq \Blambda$, and they are compatible, in principle, with the existence of null phases $\S_0$ and signature-changing sets $S$.

Next, we discuss all these different possibilities.
\subsection{Constant-signature branes or shells in AdS$_5$}
These cases are characterized by having only one of the possible phases, and thus $S=\emptyset$. The relevant physical case is the Lorentzian one, that is, when $\S=\S_L$. Then, relations (\ref{eq:cont_L}-\ref{eq:Friedmann_brane}) hold on the entire $\S$. The other two cases $\S=\S_E$ and $\S=\S_0$ can also be treated in the formalism, but they have no direct physical interpretation apart from possible topological defects.

\subsubsection{$Z_2$-mirror Lorentzian branes}
For Lorentzian branes $\S=\S_L$ with $Z_2$-mirror symmetry one only has to take 
$$\lambda=\Blambda$$
and $\varepsilon=-1$. The latter is necessary because for a $Z_2$ matching, $t=\tilde{t}$ and $r= \tilde{r}$
and (\ref{eq:ts}) implies $\sigma= \tilde{\sigma}$. Moreover, $\epsilon =-1 $ because $\bm{N}^{+}$
must be identified to $- \bm{N}^{-}$, c.f. the discussion after (\ref{eq:l1b}).
Thus $\varepsilon = -1$ follows from (\ref{varep}).
Notice that $\varepsilon =1$
corresponds to a matching that recovers the original AdS$_5$ spacetime (in particular $\varrho=0$ in that case, as
follows from 
%From our general results, this choice of sign prevents a change of signature on $\S$, since to allow
%such a change of signature we need $\varepsilon=1$.
%Observe, in any case, that as $\lambda=\Blambda$ there is no point
%in choosing  $\varepsilon=1$ anyway, since that would imply $\rho =0$ 
(\ref{eq:Friedmann_L})).
In order to have a positive $\varrho$, we have to choose
the matching such that $\sigma\epsilon_1=-1$. The geometrical
view of different possible matchings depending on
the values of $\sigma\epsilon_1$ and $\varepsilon$
are shown in Figure \ref{fig:casos}.

For these $Z_2$-symmetric branes in AdS bulks, the big-bang singularity on the brane
is characterised by the divergence of $\varrho$ and $p$.
In the cases $k=0,1$,
since the brane is assumed to be regular ($r>0$) and Lorentzian ($N<0$) everywhere,
the only possibility is that the big-bang
coincides with the vanishing of $a$.
This big-bang is therefore
located at $r\rightarrow 0$ in the AdS bulk, which
corresponds to the centre of symmetry on the bulk.
In fact, the brane cannot be regular there,
because it is forced to be Lorentzian.
As for the cases with $k=-1$, the range for $r$ is restricted to
$r>1/\lambda$ and therefore the description of $\S$ in those coordinates
obviously fails at $a\leq1/\lambda$.

\setlength{\unitlength}{1pt}
\begin{figure*}[ht]
%\graphpaper(0,0)(450,380)
\centering
\begin{picture}(450,380)
\put(130,370){AdS}
\put(300,370){$\widetilde{\mbox{AdS}}$}
\put(85,210){$\chi=0$}
\put(250,210){$\chi=0$}
\put(170,210){$\chi=\pi$}
\put(335,210){$\chi=\pi$}
\put(120,300){$1$}
\put(170,290){$2$}
\put(290,300){$\tilde{1}$}
\put(335,290){$\tilde{2}$}
\put(90,140){$\varepsilon=1$}
\put(330,140){$\varepsilon=-1$}
\put(95,70){$\epsilon_1$}
\put(310,70){$\epsilon_1$}
\put(20,70){$1$}
\put(65,50){$\tilde{2}$}
\put(140,70){$\tilde{1}$}
\put(185,50){$2$}
\put(259,60){$2$}
\put(275,55){$\tilde{2}$}
\put(350,60){$1$}
\put(420,55){$\tilde{1}$}
\put(40,80){\rotatebox{35}{$\hat\varrho>0$}}
\put(160,80){\rotatebox{35}{$\hat\varrho<0$}}
\put(40,30){\rotatebox{-35}{$\hat\varrho<0$}}
\put(160,30){\rotatebox{-35}{$\hat\varrho>0$}}
\includegraphics[width=37pc,height=30pc]{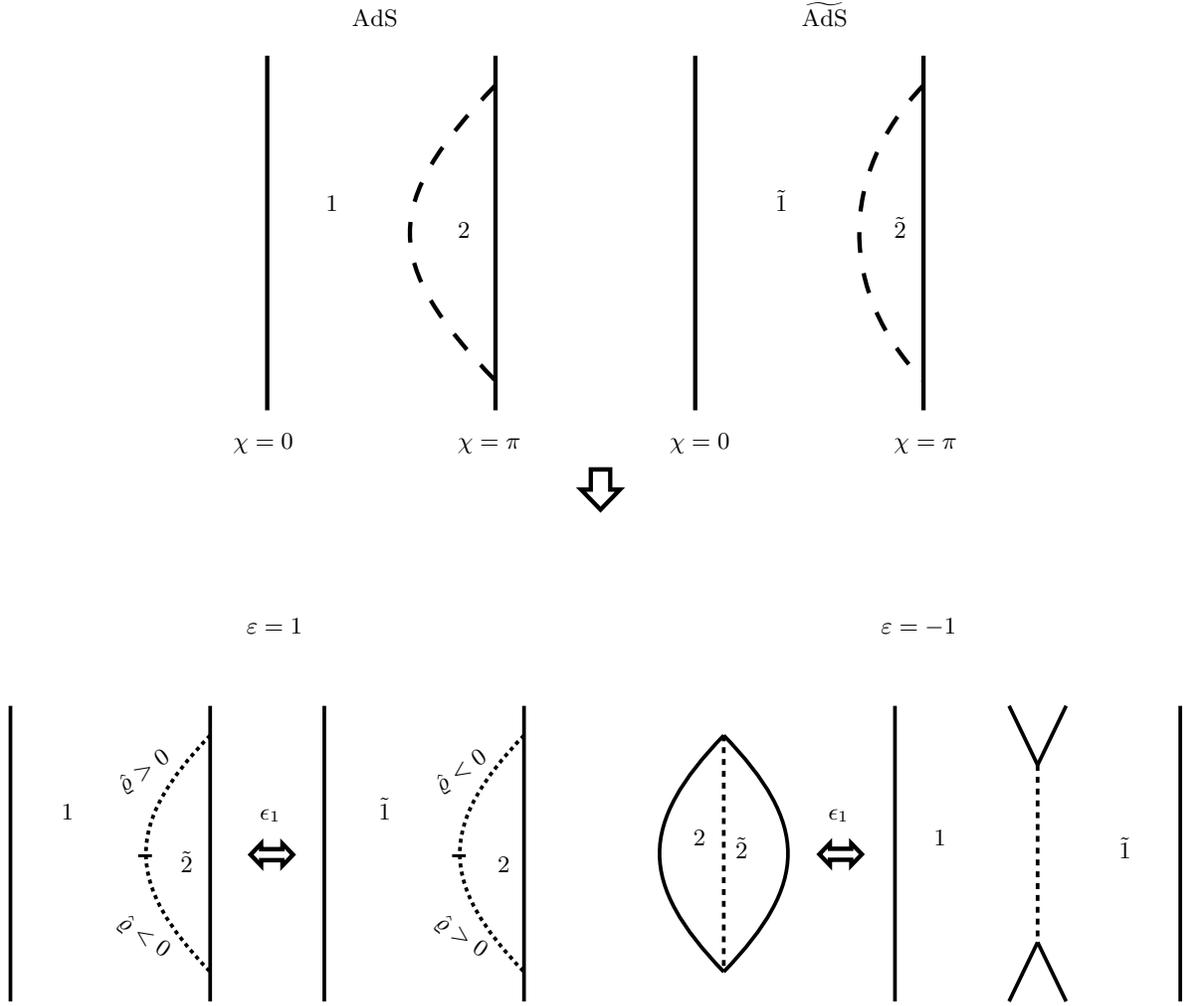}
\end{picture} 
\caption{\label{fig:casos} Four different possible matchings
between AdS and $\widetilde{\mbox{AdS}}$ in the case $k=1$
driven by the signs $\varepsilon$ and $\epsilon_1$ (fixed $\sigma$ and $\tilde{\sigma}$).
The anti de-Sitter diagrams for $k=1$ are drawn at the top. The slashed curves
represent $\S^+$ and $\S^-$, that divide
AdS and $\widetilde{\mbox{AdS}}$ into two parts, respectively.
The pairs of choices of regions (halves) to be joined
depend on the relative signs of the rigging vectors, $\varepsilon$,
and to the orientation of $\vec\ell$,
%points inwards or outwards from/to one of the halves, 
which is determined by $\epsilon_1$.
For $\varepsilon=1$ we obtain one of two possibilities on the
bottom left, which differ on the sign of $\epsilon_1$,
and for $\varepsilon=- 1$, the two on the bottom right.
}
\end{figure*}

\subsubsection{Shells, or asymmetric Lorentzian branes}
The asymmetric case is characterized by
$$\lambda \neq \Blambda .$$
Observe that then, {\em both possible signs} $\varepsilon=\pm 1$ are feasible. This has been correctly stated in \cite{AABS,AS} but,
%Nevertheless, we want to stress here the fact that, 
for unclear reasons,
only the case $\varepsilon=1$ was considered in \cite{gogb1,derdol}. By setting $\varepsilon=1$ in our formulae
%, all the expressions in
%these papers are thus recovered.
%the `shell' or `asymmetric' cosmologies the case $\varepsilon=-1$
%could have also been considered,
%but ((for some reason)) have not been found nor studied in those works.
% by choosing $\varepsilon=1$.
%In this case, 
and using the freedom in interchanging $\lambda$ and $\Blambda$
one can set $\sigma\epsilon_1=1$ without loss of generality.
This implies %$\sigma_1=-\sigma_2=1$ so that in fact
$\Blambda > \lambda$ for a positive $\varrho$.
Notice that there is an {\em upper bound} for the energy density $\varrho$ given by (\ref{limit1}). As far as we know, this upper limit had not been noticed before.
%,and so we will in the following.
%Alternatively, one could assume that $\Blambda>\lambda$
%and then play with the two values $\sigma\epsilon_1=\pm 1$.

% ((NO SE SI N'HI HA MES D'AQUESTS ARTICLES 'SHELL': si, tots els de l'equip
% d'en roy, gergely, et al. i els diuen ' asymetrics'))

%COMMENT ON WHAT IS WRONG IN DOLEZEL-DERUELLE... OTHERS?
%AIXO ES POT INDICAR A LA FIGURA...

% To have a hypersurface layer,
%and thus a brane, it suffices to assume that
%$\varepsilon\lambda-\Blambda\neq 0$.)

% Since so far we have not taken any a priori values for
% $\lambda$ and $\Blambda$, the signs $\sigma$ and $\epsilon_1$
% could be fixed now so that $\sigma\epsilon_1=1$ without loss
% of generality. The desired sign for $\varrho$ would then be
% accounted for by choosing $\lambda$ and $\Blambda$ appropriately.
% Nevertheless, these signs are used in the exposition of figure []
% for convenience. MAKE SURE OF THIS...

On the other hand, the case $\varepsilon =-1$ 
%has not been studied so far, probably due to some implicit assumptions on the %matching regions. Observe that,
requires from (\ref{eq:Friedmann_L}) that $\sigma\epsilon_1 =-1$ if $\varrho$ is to be positive. Acceptable matchings are hence possible, and both signs of $\Blambda^2-\lambda^2$ are allowed. In this case, there is a {\em lower bound} for the energy density given by (\ref{limit2}).

%\subsubsection{Null and spacelike branes}

\subsection{Signature-changing branes in AdS$_5$}
\label{tope}
It follows from Lemma \ref{uniquenessNull} that only one of the values of $\epsilon$ allows for 
a non-empty signature-changing set $S$. This was identified in (\ref{sign}) as $\epsilon =
\sign (\dot{r} \dot{\Br} \dot{t} \dot{\Bt})$. Using  (\ref{artilder}) 
and (\ref{eq:ts}) we get 
$\epsilon= \sigma\tilde{\sigma}$, so that in this subsection we must set
$$
\varepsilon = 1\, .
$$
This implies, on using (\ref{eq:Friedmann_L}), that we must necessarily require
$$\lambda \neq \Blambda$$
so that signature-changing branes must be of `asymmetric' type. This, of course, is nothing but a direct consequence of the general Corollary \ref{res:2}. Furthermore, the possible matchings for a positive $\varrho$ are identified by the necessary condition
%s $\sigma\epsilon_1=\sigma_1 =-\sigma_2$ so that 
$$\sigma\epsilon_1=\sign (\Blambda^2-\lambda^2),$$
as follows from the discussion after (\ref{limit2}).
Again, $\varrho$ is upper-bounded by (\ref{limit1}).

Nevertheless, things can behave quite differently now in comparison with the typical, purely Lorentzian,  `asymmetric' case studied above. For instance, 
%As it was found in \cite{derdol},
new types of `big-bangs' ---in the sense of the beginning of time--- can appear
at points where $a$ is not zero, $\dot a$ and $\ddot a$ are well behaved,
but where $a'$ and/or $a''$ diverge. Actually, that
happens precisely at the signature-changing set $S$
due to the vanishing of $N$ there.
%As we shall see in a moment, this kind of big-bang
%would naturally correspond to a change of signature on the brane.
This type of behaviour simply cannot be found in pure Lorentzian brane cosmologies, be them $Z_2$-symmetric or asymmetric.

Moreover, in the signature-changing case one can further prove 
that (\ref{eq:Friedmann_L}), or its consequence the modified Friedmann relation 
(\ref{eq:Friedmann_brane}), allows us to
avoid the presence of truly singular big-bangs {\em even from the
point of view of the observers in the brane}. To show this, we first note 
that $$\dot a|_S\neq 0$$ as otherwise, since $N=0$ on $S$,
from (\ref{eq:ts}) we would have that $\dot t|_S=\dot{\Bt}|_S=0$,
which we know is impossible on $S$, cf. Sect. \ref{signs}. Thus, from
%and (\ref{sigmas}) would then imply $\dot r|_S=\dot{\Br}|_S=0$,
%which would lead to a set $S$ of co-dimension three,
%with only two dimensions,
%against our assumptions. As a result, and due to
the definition of $a'$ we have the following
\begin{lemma}
On a signature-changing brane with $\S_L\neq \emptyset$, $a'$ diverges
necessarily when approaching the signature-changing set $S\cap\overline{\S_L}$. 
Hence, $H$ also diverges there. \finn
\end{lemma}

Now, since $H$ is unbounded when approaching $S\cap\overline{\S_L}$,
(\ref{eq:Friedmann_L}) easily implies that
$\varrho$ vanishes there:
\begin{equation}
\lim_{x\rightarrow S\cap\overline{\S_L}} \varrho =\lim_{x\rightarrow S\cap\overline{\S_L}} \frac{3\sigma\epsilon_1}{2\kappa^2_5}
  \frac{\Blambda^2-\lambda^2}{|H|}=0.
\label{eq:lim_rho}
\end{equation}
%Note however that $\varrho '$ will diverge at $S\cap\overline{\S_L}$, as follows from (\ref{eq:cont_L}).

Collecting the results we have thus proven the following:
\begin{theorem}
\label{res:rho_finite}
In a signature-changing brane produced
by joining AdS$_5$ and $\widetilde{\mbox{AdS}}_5$ preserving
the spatial symmetries,
the total energy density $\varrho$ on the Lorentzian part $\S_L$ of the brane is bounded above by (\ref{limit1}) and
vanishes at the set of signature-changing points $S\cap\overline{\S_L}$.\finn
\end{theorem}

As a remark, observe that linear equation of states of type $p=\gamma \varrho$ with constant $\gamma$ are not allowed in this signature-changing case, for this would imply from (\ref{eq:cont_L}) that $\varrho a^{3(1+\gamma)}=$const, which is not compatible with the vanishing of $\varrho$ at $S\cap\overline{\S_L}$ (where $a$ must be finite.) Nevertheless, general linear equations of state of type $p=p_{0} +\gamma \varrho$ are possible, as (\ref{eq:cont_L}) gives now 
$(p_{0} +(1+\gamma)\varrho)a^{3(1+\gamma)}=$const., which has no problems at $S\cap\overline{\S_L}$. Observe, however, that this particular equation of state implies clearly that $p$ must be finite at $S\cap\overline{\S_L}$.

%and therefore $\dot N$ should also vanish there.
%at $S\cap\overline{\S_L}$
%for that kind of models.
%More general barotropic equations of state are also possible.

Thus, to study the behaviour of $p$ close to the change of signature
we use (\ref{eq:p_maca}) for $\varepsilon=1$, together with (\ref{eq:lim_rho}),
to obtain the following limit
%\mnote{\marc Coefficients in last formula changed. Please check}
\begin{eqnarray*}
&&\lim_{x\rightarrow S\cap\overline{\S_L}}   p\,\, =
\lim_{x\rightarrow S\cap\overline{\S_L}}  \varrho\left(\frac{H'}{3H^2}-1\right)=\\
%  =\varrho|_S\left( \left.\frac{a a''}{a'^2}\right|_S
%    -\frac{4}{3}\right)\\
&&  = \lim_{x\rightarrow S\cap\overline{\S_L}} \frac{3 \sigma\epsilon_1}{2\kappa^2_5}
  \frac{\Blambda^2-\lambda^2}{|H|}
  \left( \frac{H'}{3H^2}-1\right)\\
&&  = \lim_{x\rightarrow S\cap\overline{\S_L}}  \frac{\sigma\epsilon_1}{2\kappa^2_5}
  (\Blambda^2-\lambda^2) \sqrt{-N}\frac{a^2}{\dot a |\dot a|}
  \left(\frac{\ddot a}{\dot a}-\frac{\dot N}{2N}
      -\frac{4 \dot a}{a}\right).
\end{eqnarray*}
Thus, the actual value of this limit
will depend on the particular choice of the function $N(\xi)$: for regular branes, $p$ diverges at $S\cap\overline{\S_L}$ if $\dot N|_S\neq 0$, while $p$ remains regular if $\dot N= 0$. Observe that changing the function $N(\xi)$ does not necessarily mean a change of cosmological model. One can obtain the same {\em Lorentzian cosmological model} ---i.e. the function $a(T)$--- starting from {\em different functions} $N(\xi)$ as long as $a(\xi)$ is changed accordingly. 

Therefore, by choosing appropriately the hypersurface $\Sigma$ in AdS$_{5}$,
%by fine-tuning the available functions, such as $N(\xi)$, 
{\em signature-changing branes such that both $\varrho$ and $p$ remain finite and well-behaved everywhere on $\overline{\S_L}$ are feasible}. Recall that $\varrho$ always vanishes at the change of signature.
% and $p$ can be regular or unbounded there.

We would like to stress that this conclusion and theorem \ref{res:rho_finite} are very satisfactory results: the Hubble parameter $H$ ---an observable quantity--- diverges when approaching the change of signature, yet the whole geometrical structure remains unhurt and the relevant physical quantities, such as $\varrho$ and $p$, are regular there.

Fully explicit examples of signature changing branes, with particular known functions $a(\xi)$ and $N(\xi)$, were presented in \cite{letter}. We refer to this letter for some discussion and extra comments of physical interest. 

\subsubsection{The Lorentzian phase $\S_L$ considered as a classical spacetime in General Relativity}
An observer living on the Lorentzian part $\S_L$ of the brane might interpret, in principle, %that the energy
%density ${\varrho}$ and pressure ${p}$
%blow up when approaching a change of signature. In other words,
%from his/her point of view, it might seem 
that a change of signature would correspond to a singularity
in a RW spacetime. If this ``singularity'' is in the past, it could represent a
big-bang from the inner point of view of $\S_L$. We would like to discuss this now in detail.

%Nevertheless,
%we are going to see in what follows some key
%differences between this part of the set $S$ and the traditional big-bang singularities, 
%such as for instance that in the present case ${\varrho}$
%is regular at the signature change, and $p$ can be also regular
%in some cases.

To begin with, it may seem contrary to our physical intuition that $\varrho\rightarrow 0$ at the signature-changing set $S\cap\overline{\S_L}$, which plays the role of such a `singularity' from the inner point of view of the Lorentzian phase $\S_L$. The meaning of this is that the total energy density $\varrho$ (the matter and radiation energy density plus the brane tension) `starts' at $S\cap\overline{\S_L}$, which is the origin of time in $\S_{L}$, with a vanishing value which increases from then on but is always bounded by (\ref{limit1}). We must remark, however, that the usual 4-dimensional Einstein equations
do {\em not} apply anywhere on $\S$, and that $\varrho$ and $p$ are (normalized) quantities associated to the singular part $\tau_{\mu\nu}$, with support on $\S$, of the energy-momentum distribution $\ttm_{\mu\nu}$.

But what would an uninformed scientist, confined to live within $\S_L$, interpret about these facts? If this scientist believes that General Relativity (GR) is the correct theory describing the universe (i.e., $\S_L$ for him/her), he/she would rather try to compute the eigenvalues of the {\em Einstein tensor} within the brane, that is to say, the Einstein tensor of the first fundamental form $h_{ab}$ of $\S_L$. The eigenvalues of this tensor are, of course,
\begin{eqnarray}
8\pi G\, \varrho^{(GR)}+\lambdafour&=&\frac{3}{a^2}(a'^2+k)=\frac{3}{a^2}\left(-\frac{\dot{a}^2}{N}+k\right), \label{eq:4d_Friedmann}\\
8\pi G \, p^{(GR)}-\lambdafour&=&-2\frac{a''}{a}-\frac{1}{a^2}(a'^2+k)=\nonumber\\
\frac{1}{N}\left(2\frac{\ddot a}{a}-\frac{\dot N}{N}\frac{\dot a}{a}\right)&+&
\frac{1}{a^2}\left(\frac{\dot{a}^2}{N}-k\right)\nonumber
\end{eqnarray}
which obviously diverge at the `singularity' placed on the signature-changing set $S\cap \overline{\S_L}$.  Here, $\lambdafour$ is the GR cosmological constant as computed by that scientist.

More importantly, let us stress the fact that there is a relation between these GR quantities and the actual energy density and pressure on the brane according to the real 5-dimensional field equations. For instance, from (\ref{eq:Friedmann_L}) we derive
\begin{eqnarray*}
\frac{\kappa^2_5}{3}\varrho=\sigma\epsilon_1
\left\{\sqrt{\frac{8\pi G}{3}\varrho^{(GR)}+\frac{\lambdafour}{3}+\Blambda^2 }-\right.\\
\left.\sqrt{\frac{8\pi G}{3}\varrho^{(GR)}+\frac{\lambdafour}{3}+\lambda^2 } \right\}
\end{eqnarray*}
while (\ref{sigma1}) and (\ref{sigma2}) give the inverse relations
$$
8\pi G\,  \varrho^{(GR)}+\lambdafour+3\Blambda^2 =
\frac{27}{4\kappa_5^4\varrho^2}\left(\Blambda^2-\lambda^2+\frac{\kappa_5^4}{9}\varrho^2\right)^2
$$
and the one obtained by interchanging $\Blambda \leftrightarrow \lambda$.

These formulas patently show that the GR `singularity' where 
$\varrho^{(GR)}\rightarrow \infty$, which corresponds to the signature change, is simply a manifestation of the fact that the proper energy density on the Lorentzian phase of the brane actually vanishes there.

\subsection{Recovering the Friedmann equation at different limits:
effective 4-dimensional fundamental constants}
A well-known fact in the $Z_2$-symmetric brane cosmologies,
as well as in the asymmetric ``shell'' cosmologies,
is that the usual Friedmann equation (\ref{eq:4d_Friedmann}) of General Relativity
%\begin{equation}
%  \label{eq:4d_Friedmann}
%  H^2+\frac{k}{a^2}=\frac{8\pi G}{3}\varrho^{(GR)} + \frac{\lambdafour}{3},
%\end{equation}
%where $\lambdafour$ is the four-dimensional
%cosmological constant and $\varrho_m$ the `matter' energy density,
% and $\rho_m$ refers to the 
can be recovered from the equation on the brane (\ref{eq:Friedmann_L})
at the limit when the matter density is small compared to $\Lambda$, once a non-vanishing tension
$\tension$ has been introduced in an appropriate manner. This limit is,
in fact, the one used to recover
the full 4-dimensional Einstein equations in GR, see \cite{SMS,roy_review},
and to relate $\Lambda_5$ and $\tension$ with effective
4-dimensional gravitational and cosmological constants (see (\ref{eq:usual_fc}) below). 

Nevertheless, that limit relies on the existence of a non-vanishing tension.
In the present case there is another
%, one can take {\em at least} another
limit, both natural and convenient, for which no tension is needed.
Such limit corresponds to large values of $a$
%(or better $\lambda a$ and $\Blambda a$,
%while also keeping a finite $a'$, so that $H/\lambda^2$ is small).
while keeping a finite $a'$, so that $H^2+k/a^2$ is small.
Another  characterization of this limit is that 
$8 \pi G \varrho^{(GR)}+\lambdafour$ is small. One
appropriate dimensionless quantity to perform rigorously this limit is 
$$
\frac{H^2+k/a^2}{\lambda^2} \approx \frac{8 \pi G \varrho^{(GR)}+\lambdafour}{\Lambda_{5}},
$$ 
where $\approx$ stands for equality except for a constant of order one. 
$\Blambda$ or $\widetilde{\Lambda_{5}}$ could also be used to define the dimensionless parameter.

Also worth mentioning is the fact that in many papers
(see \cite{SMS,roy_review,gergely1})
the limits of the modified Friedmann equation have been
taken starting from the quadratic equation (\ref{eq:Friedmann_brane}),
instead of the original (\ref{eq:Friedmann_L}) which contains more information,
thus missing the meaning of the signs $\varepsilon$ and $\epsilon_1\sigma$.
An exception is \cite{derdol} where the authors considered
equation (\ref{eq:Friedmann_L}), but as mentioned before not all the possible signs
were taken into account. Therefore, 
for the sake of completeness, 
%and since in previous works some cases
%depending on $\varepsilon$ and $\epsilon_1\sigma$
%were not taken into consideration, 
let us derive the limits keeping those signs free.

%\mnote{\marc Title changed. I don't think $\rho^{GR}$ must be small. Only the combination with
%$\Lambda_4$, which makes the title long}
\subsubsection{Large values of $a$ with small values of $H$}
Let us start by considering the limit for large values of $a$
while keeping $H$ small. % (and the same with tildes).
Equation (\ref{eq:Friedmann_L}) for $n=5$ can be approximated to
\[
 \sigma\epsilon_1 \frac{\kappa_5^2}{3} \varrho= \varepsilon \Blambda -\lambda +
\frac{1}{2}\frac{a'^2 + k}{a^2}
\left(\frac{\varepsilon}{\Blambda}-\frac{1}{\lambda}\right)
+ O(a^{-4}).
\]
Since $\varepsilon\lambda-\Blambda\neq 0$
in order to have a brane or shell at all,
this expression can be rearranged as
\begin{equation}
  \label{eq:limit_friedmann}
 3\left( \frac{a'^2}{a^2}+\frac{k}{a^2}\right)=\sigma\epsilon_1 \kappa_5^2
  \frac{2\Blambda\lambda}
  {\varepsilon\lambda-\Blambda}\varrho
  +6\varepsilon\Blambda\lambda + O(a^{-4}).
\end{equation}
%whenever $\epsilon\lambda-\Blambda\neq 0$.
Let us consider now the tension of the brane
as a contributing part of $\ttt_{ab}$, so that (\ref{eq:tensionandmatter})
holds. Then, $\varrho$ and $p$ decompose as
$\varrho=\varrho_m + \tension$ and
$p=p_m - \tension$, where $\varrho_m$ and $p_m$ correspond to
$\ttt^m_{ab}$.
Using this together with (\ref{eq:4d_Friedmann}) in
(\ref{eq:limit_friedmann}), we derive
$$
8\pi G\, \varrho^{(GR)}+\lambdafour=\sigma\epsilon_1 \kappa_5^2
  \frac{2\Blambda\lambda}
  {\varepsilon\lambda-\Blambda}(\varrho_{m}+\Lambda)
  +6\varepsilon\Blambda\lambda + O(a^{-4}).
$$
There are many ways to interpret this relation.
In principle,
it simply determines the value of $\varrho^{(GR)}$
in terms of $\rho_{m},\Lambda$ and the constants $G,\kappa_{5},\lambda,\Blambda$ and $\Lambda_{4}$. It seems natural, however, to identify the constant terms at both sides of this relation, and therefore the remaining terms too.
Identifying 
%By doing this such that 
\begin{equation}
\varrho^{(GR)}\longleftrightarrow \varrho_{m}\label{identify}
\end{equation}
we obtain the following relations between the fundamental constants:
\begin{eqnarray}
  8 \pi G &=& \sigma\epsilon_1\kappa_5^2\frac{2\Blambda\lambda }
  {\varepsilon\lambda-\Blambda}\, ,\label{eq:G_1}\\
  \lambdafour&=&\sigma\epsilon_1\kappa_5^2 \frac{2\Blambda\lambda }
  {\varepsilon\lambda-\Blambda}
  \tension + 6 \varepsilon \Blambda\lambda\, .\label{eq:cc4_1}
\end{eqnarray}
As far as we are aware, these relations were previously unknown.

Relations (\ref{eq:G_1}) and (\ref{eq:cc4_1}) can be particularised to the case of
$Z_2$-symmetric Lorentzian branes, for which $\lambda=\Blambda$ and  $\varepsilon=-1$, so that
\[
  8 \pi G =-\sigma\epsilon_1 \kappa_5^2 \lambda,~~~
  \lambdafour= -\sigma\epsilon_1 \kappa_5^2\lambda\tension-6\lambda^2.
\]
In view that we need $\sigma\epsilon_1=-1$ for
a positive gravitational constant %and a positive $\varrho$,
these equations can be rearranged as
\begin{equation}
\label{eq:new_fc}
8 \pi G =\frac{\kappa_5^4}{6}
\left(\tension -\frac{\lambdafour}{8\pi G}\right),~~~
\lambdafour=8\pi G\tension + \Lambda_5.
\end{equation}
These expressions %, which we have not found in the literature,
differ from the ones usually obtained in the literature, involving a different limit 
---given by (\ref{eq:usual_fc}) below---, and seem to be new.
%we have not been able to find them in the literature.

\subsubsection{Small values of $\varrho_{m}/\tension$}
As for the usual limit $\varrho_m/\tension\to 0$ with a non-vanishing
$\tension$ it is convenient to start from the quadratic
equation (\ref{eq:Friedmann_brane}). Using $\varrho=\varrho_m+\tension$
and defining $\beta\equiv 3 \lambda/(\tension \kappa_5^2)$ and
$\tilde\beta\equiv 3 \Blambda/(\tension \kappa_5^2)$,
it can be expressed as
\begin{eqnarray*}
  \frac{a'^2}{a^2}+\frac{k}{a^2}&=&
  \frac{\kappa_5^4}{36}\tension^2
  \left[1-2(\tilde\beta^2+\beta^2)+(\tilde\beta^2-\beta^2)^2\right]\\
  &+&\frac{\kappa_5^4}{18}\tension^2\left[1-(\tilde\beta^2-\beta^2)^2\right]
  \frac{\varrho_m}{\tension}+O[(\varrho_m/\tension)^2].
\end{eqnarray*}
Comparing with (\ref{eq:4d_Friedmann}) we get 
\begin{eqnarray*}
8\pi G\, \varrho^{(GR)}&+&\lambdafour= \frac{\kappa_5^4}{12}\tension^2
  \left[1-2(\tilde\beta^2+\beta^2)+(\tilde\beta^2-\beta^2)^2\right]\\
  &+&\frac{\kappa_5^4}{6}\tension^2\left[1-(\tilde\beta^2-\beta^2)^2\right]
  \frac{\varrho_m}{\tension}+O[(\varrho_m/\tension)^2]
\end{eqnarray*}
which, as before, provides an expression for $\varrho^{(GR)}$ and can be resolved in many different ways. Using again the natural identification (\ref{identify}),
a different set of
relations for the effective fundamental constants is obtained:
%\mnote{\marc Several $\kappa_5^2$ changes to $\kappa_5^4$. Please check}
\begin{eqnarray}
  8 \pi G &=&\frac{1}{6}\kappa_5^4\tension
  \left[1-\frac{81}{\kappa_5^8 \tension^4}(\Blambda^2-\lambda^2)^2\right],
  \label{eq:G_2}\\
  \lambdafour&=& \frac{1}{12}\kappa_5^4\tension^2
  \left[1-\frac{18}{\kappa_5^4 \tension^2}(\Blambda^2+\lambda^2)\right.
  \nonumber\\
    &&\left.+\frac{81}{\kappa_5^8 \tension^4}(\Blambda^2-\lambda^2)^2\right].
  \label{eq:cc4_2}
\end{eqnarray}
For the particular $Z_2$-symmetric branes, for which
$\lambda=\Blambda$, and recalling that $\lambda^2=-\Lambda_5/6$,
these two relations simplify to
\begin{equation}
  8 \pi G =\frac{1}{6}\kappa_5^4\tension,~~~
  \lambdafour= \frac{1}{2}(8\pi G\tension+\Lambda_5),
  \label{eq:usual_fc}
\end{equation}
which correspond to the usual relations found
in the literature \cite{roy_review}.

\subsubsection{Relationship between the two limits}
We have seen that the usual relations (\ref{eq:usual_fc}) are not unique,
since they depend crucially on the kind of limit taken. Another limit of physical interest, with no need of a tension $\Lambda$, leads for instance to the alternative relations (\ref{eq:new_fc}).
These two sets (\ref{eq:usual_fc}) and (\ref{eq:new_fc})
only coincide when one demands a vanishing effective four-dimensional
cosmological constant, this is, if the tension of the brane
is fine tuned in order to have $\lambdafour=0$.
In that case both sets contain the same information,
given by $\kappa_5^4 \tension^2 /6=-\Lambda_5$ (the fine-tuning
of the tension)
and $8\pi G=\kappa_5^4 \tension /6$.
%This is the reason why in reference \cite{derdol}, although
This was to be expected, because the limit at large $a$ implies
%that $\varrho$ tends to zero, so
that  $\varrho^{(GR)}$ tends to the constant
$-\lambdafour/8\pi G$, and therefore if (and only if) $\lambdafour$
vanishes then the limit $\varrho_m/\tension\to 0$ is recovered by means of the identification (\ref{identify}).

It is worth mentioning here that in reference \cite{derdol}, despite the use
of the limit of a particular case of expression (\ref{eq:cont_L}),
the relations found for the fundamental constants are
the usual ones (\ref{eq:usual_fc}) precisely because it was assumed that
$\lambdafour=0$.

\begin{acknowledgments}
MM was supported by the projects
FIS2006-05319 of the Spanish CICyT,
SA010CO of the Junta de Castilla y Le\'on and
P06-FQM-01951 of the Junta de Andaluc\'{\i}a.
JMMS thanks financial support under grants FIS2004-01626 of the Spanish CICyT and GIU06/37 of the University of the Basque Country (UPV).
RV is funded by the Basque Government Ref. BFI05.335 and
thanks financial support from project GIU06/37 (UPV).
\end{acknowledgments}

%\section{Discussion LA FEM?}
%Insist on FREEDOM IN CHOOSING $a$ and $N$?

%Do we insist on the two different sets of relations for
%the fundamental constants?

\section{Appendix}
\label{app:Hs}
The aim of this appendix is to present the intermediate steps leading
from the expressions (\ref{exprhh}) and (\ref{exprhhxixi}) involving $L$
and $\alpha$ to the final result (\ref{finalhh}) and
(\ref{finalhhxixi}) which is independent of $L$ and $\alpha$ and
symmetric under the interchange of $\N^{+}$ by $\N^{-}$.

Regarding $[\hh]$, it  turns out to be convenient to work with
$ \ff [ \hh ]$. Directly from their definitions (\ref{ffpm}), (\ref{hhpm})
we have, using $C\eqq \BC \equiv a(\xi)$,
\begin{eqnarray}
  \ff [ \hh ]  &=& \ff^{+} \hh^{+} - \ff^{-} \hh^{-} = 
  \nonumber \\
  &&a \epsilon_1 
  \left\{ \dot{t} \frac{C_{,r}}{B^2} \left (-A^2 \dot{t}^2
      + B^2 \dot{r}^2 \right )
  \right.
  \nonumber \\
  &&\left .\left .
      - L \dot{\Bt} \frac{\BC_{,\Br}}{\BB^2} 
      \left (- \BA^2 \dot{\Bt}^2 + \BB^2 \dot{\Br}^2 \right )
    \right\} \right |_{\S}.
  \label{hh2}
\end{eqnarray}
Taking now the derivative of $a(\xi)$ we get
$  \dot{a} = \left. C_{,r} \dot{r} \right |_{\S^{+}} = 
\left . \BC_{,\Br} \dot{\Br}  \right |_{\S^{-}}$, which
allows us to build the following chain of equalities
\begin{eqnarray*}
  &&C_{,r} \dot{r}^2 \dot{t} - \BC_{,\Br} L \dot{\Br}^2 \dot{\Bt}
  = \dot{a} \left (\dot{r} \dot{t} - L \dot{\Br} \dot{\Bt} \right )
  \\
  &&= \dot{a} \left ( \alpha^2 L \dot{\Bt} \dot{\Br} - \dot{t} \dot{r}
  \right ) = \left .\BC_{,\Br}  L \alpha^2\dot{\Bt} \dot{\Br}^2 -
    C_{,r} \dot{t} \dot{r}^2  \right |_{\S},
\end{eqnarray*}
where in the second equality we used (\ref{eq:l1a}).
Substituting now the term
$C_{,r} \dot{r}^2 \dot{t} 
-  \BC_{,\Br} L \dot{\Br}^2 \dot{\Bt}$ appearing in (\ref{hh2}) by this
expression, we find
\begin{eqnarray*}
  \ff [ \hh ]  &=& a \epsilon_1 
  \left\{ - C_{,r} \dot{t} \frac{A}{B}
    \left (\frac{A}{B} \dot{t}^2 + \frac{B}{A}\dot{r}^2 
    \right )\right.
  \\
  &&\left.\left.+ \BC_{,\Br} \dot{\Bt} L \frac{\BA}{\BB}
      \left ( \frac{\BA}{\BB}
        \dot{\Bt}^2 + \alpha^2 \frac{\BB}{\BA}\dot{\Br}^2 \right )
    \right\} \right |_{\S}.
\end{eqnarray*}
It only remains to use (\ref{eq:l1b}) in the two terms
in parenthesis in order to get the final result
(\ref{finalhh}).

Let us now rewrite (\ref{exprhhxixi}) in a symmetric manner. To that aim,
it is convenient to consider $[\hh_{\xi\xi}] \dot{t} \dot{\Bt}$ and try to
get common factors $\ell_{+}^{\alpha}\nany^{+}_{\alpha}$ as we did before. 
Rearranging terms in (\ref{exprhhxixi}) yields
\begin{widetext}
\begin{eqnarray}
\epsilon_1\,[\hh_{\xi\xi}] \dot{t} \dot{\Bt} &\eqq&
\dot{\Bt} \left\{
- \frac{A_{,r}}{A} \dot{r}^2 \dot{t}^2 + \frac{B B_{,r}}{A^2}
\dot{r}^4 - \dot{r} \dot{t} \ddot{t} - \dot{t}^2 \ddot{r} 
- \left (\frac{A_{,r}}{B} \dot{t}^2 + \frac{B_{,r}}{A}
\dot{r}^2 \right ) \left ( \frac{A}{B} \dot{t}^2 + \frac{B}{A} \dot{r}^2
\right ) \right\}
\nonumber \\
&&-\dot{t} L \left\{
- \alpha^2 
\frac{\BA_{,\Br}}{\BA} \dot{\Br}^2 \dot{\Bt}^2 + \alpha^2
 \frac{\BB \BB_{,\Br}}{\BA^2}
\dot{\Br}^4 -  \alpha^2 \dot{\Br} \dot{\Bt} \ddot{\Bt} -
 \dot{\Bt}^2 \ddot{\Br} -
\left (\frac{\BA_{,\Br}}{\BB} \dot{\Bt}^2 + 
\frac{\BB_{,\Br}}{\BA}
\dot{\Br}^2 \right ) \left ( \frac{\BA}{\BB} \dot{\Bt}^2 + \alpha^2
\frac{\BB}{\BA} \dot{\Br}^2
\right ) 
\right\}.
\label{hhxixi2}
\end{eqnarray}
%\end{widetext}
Now, evaluating $\dot{\ff}^{\pm}$  allows us to write the identities
\begin{eqnarray*}
 - \frac{A_{,r}}{A} \dot{r}^2 \dot{t}^2 + \frac{B B_{,r}}{A^2}
\dot{r}^4 - \dot{r} \dot{t} \ddot{t} - \dot{t}^2 \ddot{r} 
%\\
%&&\qquad\qquad
&\eqq& \frac{1}{2 A^2} \dot{r} \dot{\ff} - \frac{B}{A} \ddot{r}
\left ( \frac{A}{B} \dot{t}^2 + \frac{B}{A} \dot{r}^2 \right ) ,
\\
%\end{eqnarray*}
%and
%\begin{eqnarray*}
- \alpha^2 
\frac{\BA_{,\Br}}{\BA} \dot{\Br}^2 \dot{\Bt}^2 +  \alpha^2
 \frac{\BB \BB_{,\Br}}{\BA^2}
\dot{\Br}^4 -  \alpha^2 \dot{\Br} \dot{\Bt} \ddot{\Bt} -
 \dot{\Bt}^2 \ddot{\Br} 
%\\
%&&
%\qquad\qquad
&\eqq&
 \frac{\alpha^2 }{2 \BA^2} \dot{\Br} \dot{\ff} -  \frac{\BB}{\BA}
\ddot{\Br}
\left ( \frac{\BA}{\BB} \dot{\Bt}^2 + \alpha^2 \frac{\BB}{\BA} 
\dot{\Br}^2 \right ),
\end{eqnarray*}
which substituted in (\ref{hhxixi2}) yields
%\begin{widetext}
\begin{eqnarray}
\epsilon_1\,[\hh_{\xi\xi}] \dot{t} \dot{\Bt} &\eqq&
\dot{\Bt} \left\{
\frac{1}{2 A^2} \dot{r} \dot{\ff} - \left (
\frac{B}{A} \ddot{r} + \frac{A_{,r}}{B} \dot{t}^2 + \frac{B_{,r}}{A}
\dot{r}^2 \right ) \left ( \frac{A}{B} \dot{t}^2 + \frac{B}{A} \dot{r}^2
\right ) \right\}
\nonumber \\
&&-\dot{t} L \left\{ 
\frac{\alpha^2 }{2 \BA^2} \dot{\Br} \dot{\ff} - \left (  \frac{\BB}{\BA}
\ddot{\Br} +
\frac{\BA_{,\Br}}{\BB} \dot{\Bt}^2 + 
\frac{\BB_{,\Br}}{\BA}
\dot{\Br}^2 \right ) \left ( \frac{\BA}{\BB} \dot{\Bt}^2 + \alpha^2
\frac{\BB}{\BA} \dot{\Br}^2
\right ) 
\right\}. \label{hhxixi3}
\end{eqnarray}
%\end{widetext}
In this expression the only terms that require extra treatment are
$A^{-2} \dot{r} \dot{\Bt} - L \alpha^2  \BA^{-2} \dot{\Br} \dot{t} $.
Multiplying the first summand by $\ff^{+}$ and the second by $\ff^{-}$
we get, after adding zero in the form
of $2 \dot{\Bt} \dot{r} \dot{t}^2 - L (1+\alpha^2) \dot{\Br} \dot{\Bt}^2
\dot{t}= 0 $, see (\ref{eq:l1a}),
\begin{eqnarray*}
  %&&
  \ff \left ( \frac{1}{A^2} \dot{r} \dot{\Bt} -
    L \alpha^2  \frac{1}{\BA^{2}} \dot{\Br} \dot{t} \right)  \eqq
  %\\
  %&&
  \frac{B}{A} \dot{\Bt} \dot{r} \left (
    \frac{A}{B} \dot{t}^2 + \frac{B}{A} \dot{r}^2 \right) - L \frac{\BB}{\BA}
  \dot{t} \dot{\Br} \left (  \frac{\BA}{\BB} \dot{\Bt}^2 +
    \alpha^2  \frac{\BB}{\BA} \dot{\Br}^2 \right) .
\end{eqnarray*}
Inserting this into (\ref{hhxixi3}) we finally find
%\begin{widetext}
\begin{eqnarray*}
\epsilon_1\,[\hh_{\xi\xi}] \dot{t} \dot{\Bt} &\eqq&
\dot{\Bt} \left ( \frac{\dot{N}}{2N} \frac{B}{A} \dot{r} -
\frac{B}{A} \ddot{r} - \frac{A_{,r}}{B} \dot{t}^2 - \frac{B_{,r}}{A}
\dot{r}^2 \right ) \left ( \frac{A}{B} \dot{t}^2 + \frac{B}{A} \dot{r}^2
\right ) 
- \nonumber \\
&&\dot{t} L \left ( \frac{\dot{N}}{2 N} \frac{\BB}{\BA} \dot{\Br}
- \frac{\BB}{\BA}
\ddot{\Br} -
\frac{\BA_{,\Br}}{\BB} \dot{\Bt}^2 - 
\frac{\BB_{,\Br}}{\BA}
\dot{\Br}^2 \right ) \left ( \frac{\BA}{\BB} \dot{\Bt}^2 + \alpha^2
\frac{\BB}{\BA} \dot{\Br}^2
\right ), 
\end{eqnarray*}
\end{widetext}
which  becomes exactly (\ref{finalhhxixi}) after using (\ref{eq:l1b}).


\begin{thebibliography}{999}
\bibitem{letter} M. Mars, J.M.M. Senovilla, R. Vera \Journal{\PRL}{86}
{4219}{-4222}{2001}
{Signature Change on the Brane}

\bibitem{arkani} N. Arkani-Hamed, S. Dimopoulos, G. Dvali \Journal{\PLB}{429}
{263}{-272}{1998}
{The Hierarchy Problem and New Dimensions at a Millimeter}

\bibitem{r-sI} L. Randall, R. Sundrum, \Journal{\PRL}{83}
{3370}{-3373}{1999}
{A Large Mass Hierarchy from a Small Extra Dimension}

\bibitem{r-sII} L. Randall, R. Sundrum, \Journal{\PRL}{83}
{4690}{-4693}{1999}
{An Alternative to Compactification}

\bibitem{barcelo} C. Barcel\'o, M. Visser \Journal{\PLB}{482}
{183}{-194}{2000}
{Living on the edge: cosmology on the boundary of anti-de Sitter space} 

\bibitem{hig-dim} J.M. Overduin, P.S. Wesson \Journal{\PREP}{283}
{303}{-380}{1997}
{Kaluza-Klein Gravity} 

\bibitem{GI} G.W. Gibbons, A. Ishibashi, \Journal{\CQG}{21}{2919}{-2936}{2004}{Topology and Signature Changes in Braneworlds}

\bibitem{AABS} S. Ansoldi, A. Aurilia, R. Balbinot, E. Spallucci \Journal{\CQG}{14}{2727}{-2755}{1997}{Classical and quantum shell dynamics, and vacuum decay}

\bibitem{SMS} T. Shiromizu, K. Maeda, M. Sasaki \Journal{\PRD}{62}
{024012-1}{-6}{2000}
{The Einstein equations on the 3-brane world}

\bibitem{gogb1} M. Gogberashvili, \Journal{\EPL}{49}
{396}{-399}{2000}
{Our world as an expanding shell} 

\bibitem{derdol}
N. Deruelle, T. Dolezel \Journal{\PRD}{62}
{103502}{}{2000}
{Brane versus shell cosmologies in Einstein and Einstein-Gauss-Bonnet theories} 

\bibitem{Carter} B. Carter \Journal{\IJT}{40}
{2099}{-2130}{2001}
{Essentials of Classical Brane Dynamics} 
%for a purely classical description of branes.  

\bibitem{gergely1} L.A. Gergely \Journal{\PRD}{68}
{124011}{}{2003}
{Generalized Friedmann branes}

\bibitem{roy_review} R. Maartens (2004)
``Brane-World Gravity'', Living Rev. Relativity \textbf{7},    3.
URL (cited on 16 April 2007):http://www.livingreviews.org/lrr-2004-7

\bibitem{AS} S. Ansoldi, L. Sindoni (2007) ``Shell-mediated tunnelling between (anti-)de Sitter vacua'',  arXiv:0704.1073

\bibitem{Israel66} W. Israel \Journal{\NC}{B44}
{1}{-14}{1966}
{Singular hypersurfaces and thin shells in General Relativity};
erratum {\bf 48}, 463  (1967)

\bibitem{sing} S.W. Hawking, G.F.R. Ellis, {\it The large scale structure 
of space-time}, (Cambridge Univ. Press, Cambridge, 1973) 

\bibitem{sing2} J.M.M. Senovilla, \Journal{\GRG}{30}
{701}{-848}{1998}
{Singularity theorems and their consequences} 

\bibitem{HH} T. Hirayama, B. Holdom \Journal{\PRD}{68}{044003}{}{2003}{Can black holes have Euclidean cores?}

\bibitem{noboundaryprop} J.B. Hartle, S.W. Hawking \Journal{\PRD}{28} 
{2960}{-2975}{1983}
{Wave function of the Universe}

\bibitem{Sak} A.D. Sakharov (1984) ``Cosmological transitions with change of signature'' {\it Sov.Phys.JETP} {\bf 60} 214-218 

\bibitem{vilenkintunnel} A. Vilenkin \Journal{\PRD}{33}
{3560}{-3569}{1986}
{Boundary conditions in quantum cosmology}

\bibitem{chsig}
T. Dray, A.M. Corinne, R.W. Tucker \Journal{\GRG}{23}
{967}{-971}{1991}
{Particle production from signature change}

\bibitem{chsig2} S.A. Hayward \Journal{\CQG}{9}
{1851}{-1862}{1992}
{Signature change in general relativity}

\bibitem{ellissig}G. Ellis, A. Sumeruk, D. Coule, C. Hellaby 
\Journal{\CQG}{9}{1535}{-1554}{1992}
{Change of signature in classical relativity}

\bibitem{chsig3} M. Kriele, J. Martin \Journal{\CQG}{12}
{503}{-511}{1995}
{Black holes, cosmological singularities and change of signature}

\bibitem{DEH} T. Dray, G. Ellis, C. Hellaby \Journal{\GRG}{33}
{1041}{-1046}{2001}
{Note on Signature Change and Colombeau Theory}

\bibitem{taylor} J.P.W. Taylor 
\Journal{\CQG}{21}{3705}{-3715}{2004}{Junction conditions at a corner}

\bibitem{Germani01} C. Germani, R. Maartens \Journal{\PRD}{64}
{124010}{}{2001}
{Stars in the brane world}

\bibitem{Dadhich02} M. Govender, N. Dadhich \Journal{\PLB}{538}
{233}{-238}{2002}
{Collapsing sphere on the brane radiates}

\bibitem{DT} T. Dereli, R. W. Tucker \Journal{\CQG}{10}{365}{-373}{1993}{Signature dynamics in general relativity}

\bibitem{DR} F. Darabi, A. Rastkar, \Journal{\GRG}{38}{1355}{-1366}{2006}{A quantum cosmology and discontinuous signature changing classical solutions}

\bibitem{VJS} B. Vakili, S. Jalalzadeh, H.R. Sepangi, \Journal{\JCAP}{05}{006}{}{2005}{Classical and quantum spinor cosmology with signature change}

\bibitem{VS} B. Vakili, H.R. Sepangi, \Journal{\JCAP}{09}{008}{}{2005}{Bianchi-I classical and quantum spinor cosmology with signature change}

\bibitem{VJS2} B. Vakili, S. Jalalzadeh, H.R. Sepangi, \Journal{\ANN}{321}{2491}{-2503}{2006}{Compactification and signature transition in Kaluza-Klein spinor cosmology}

\bibitem{DDT} A. Das, A. DeBenedictis, N. Tariq \Journal{\JMP}{44}{5637}{-5655}{2003}{General solution of Einstein's spherically symmetric gravitational equations with junction conditions}

\bibitem{GGI} C. Galfard, C. Germani, A. Ishibashi \Journal{\PRD}{73}{064014}{}{2006}{Asymptotically AdS brane black holes}

\bibitem{MarsSenovilla93} M. Mars, J.M.M. Senovilla \Journal{\CQG}{10}
{1865}{-1897}{1993}
{Geometry of general hypersurfaces in spacetime: junction conditions}

\bibitem{Sha} M.S. Shapovalova (2001) ``Large fluctuations of time and change of space-time signature'' 
{\it Grav. Cosmol.} {\bf 7} 193-196 

\bibitem{HHW} K. Hashimoto, P.-M. Ho, J.E. Wang \Journal{\PRL}{90}{141601}{}{2003}{Spacelike brane actions}

\bibitem{HHNW} K. Hashimoto, P.-M. Ho, S. Nagaoka, J.E. Wang \Journal{\PRD}{68}{026007}{}{2003}{Time evolution via S-branes}

\bibitem{KP} S. Kar, S. Panda, \Journal{\JHEP}{11}{052}{}{2002}{Electromagnetic strings: complementarity between time and temperature}

\bibitem{KM} S. Kar, S. Majumdar \Journal{\IJMPA}{21}{2391}{-2401}{2006}{Scattering of noncommutative strings}

\bibitem{JAS} S. Jalalzadeh, F. Ahmadi, H.R. Sepangi \Journal{\JHEP}{08}{012}{}{2003}{Multi-dimensional classical and quantum cosmology: exact solutions, signature transition and stabilization}

\bibitem{WWV} S. Weinfurtner, A. White, M. Visser, ``Trans-Planckian physics and signature change events in Bose gas hydrodynamics'', arXiv:gr-qc/0703117

\bibitem{ClarkeDray87} C.J.S. Clarke, T. Dray \Journal{\CQG}{4}
{265}{-275}{1987}
{Junction conditions for null hypersurfaces}


\bibitem{AL} E. Aguirre-Dab\'an, J. Lafuente-L\'opez, private communication

%\bibitem{AguirreLafuente} E. Aguirre-Dab\'an, J.Lafuente-L\'opez,  
%``Transverse Riemann-Lorentz metrics with a tangent radical'', math-DG/0306153.
\bibitem{AguirreLafuente} E. Aguirre-Dab\'an and J. Lafuente-L\'opez \Journal{\em Diff. Geom. Appl.}{24}
{91}{-100}{2006}
{Transverse Riemann-Lorentz type-changing metrics with tangent radical}


\bibitem{Hirsch} M.W. Hirsch, {\it Differential Topology} (Graduate Texts in
                  Mathematics) {\bf 33}, (Springer Verlag, 1997).




% \bibitem{Feng} J.L. Feng, J. March-Russell, S. Sethi, F. Wilczek
% \Journal{\em Nucl. Phys. B}{602}
% {307}{-328}{2001}
% {Saltatory Relaxation of the Cosmological Constant}

\bibitem{mps} R. Vera \Journal{\CQG}{19}
{5249}{-5264}{2002}
{Symmetry-preserving matchings}

\bibitem{FST} F. Fayos, J.M.M. Senovilla, R. Torres \Journal{\PRD}{54}
{4862}{-4872}{1996}
{General matching of two spherically symmetric spacetimes}


% \bibitem{John} M.V. John, \Journal{\AP}{630}{667}{-674}{2005}{Cosmography, Decelerating Past, and Cosmological Models: Learning the Bayesian Way}



\end{thebibliography}
\end{document}